\documentclass[twocolumn,preprintnumbers, amssymb,amsmath,aps,floatfix,prd,nofootinbib,superscriptaddress,showpacs]{revtex4-1}
\usepackage{mathrsfs}
\usepackage{amsfonts}
\usepackage{amsmath}
\usepackage{array}
\usepackage{verbatim}
\usepackage{epsfig}
\usepackage{graphicx}

\begin{document}
\title{Lepton-jet Correlation in Deep Inelastic Scattering}

\author{Xiaohui Liu}
\affiliation{Center of Advanced Quantum Studies, Department of Physics,
Beijing Normal University, Beijing 100875, China}

\author{Felix Ringer}
\affiliation{Nuclear Science Division, Lawrence Berkeley National
Laboratory, Berkeley, CA 94720, USA}

\author{Werner Vogelsang}
\affiliation{Institute for Theoretical Physics,
                Universit\"{a}t T\"{u}bingen,
                Auf der Morgenstelle 14,
                D-72076 T\"{u}bingen, Germany}
                
\author{Feng Yuan}
\affiliation{Nuclear Science Division, Lawrence Berkeley National
Laboratory, Berkeley, CA 94720, USA}

\begin{abstract}
We study the lepton-jet correlation in deep inelastic scattering. We perform one-loop calculations for the spin averaged and transverse spin dependent differential cross sections depending on the total transverse momentum of the final state lepton and the jet. The transverse momentum dependent (TMD) factorization formalism is applied to describe the relevant observables. To show the physics reach of this process, we perform a phenomenological study for HERA kinematics and comment on an ongoing analysis of experimental data. In addition, we highlight the potential of this process to constrain small-$x$ dynamics.
\end{abstract}

\maketitle

\section{Introduction}
%{\it Introduction.}

In a recent paper~\cite{Liu:2018trl}, we have proposed the lepton-jet correlation in deep inelastic scattering (DIS) at the planned Electron-Ion Collider (EIC)~\cite{Boer:2011fh,AbelleiraFernandez:2012cc,Accardi:2012qut} as a unique probe to explore the structure of nucleons/nuclei. In this paper, we provide a detailed derivation of the formalism and perform a phenomenological study relevant for the existing jet production data in DIS from HERA~\cite{Abramowicz:2017ful,Abramowicz:2012jz,Abramowicz:2010ke}, which have been re-analyzed recently to study the lepton-jet correlation~\cite{Amilkar}. 

\begin{figure}[t]
\begin{center}
\includegraphics[width=6cm]{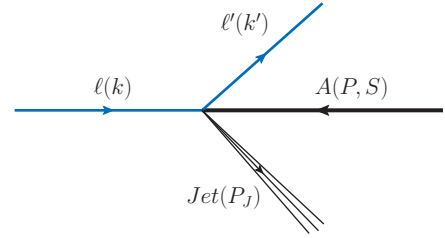}
\end{center}
\caption[*]{The lepton-jet correlation in deep-inelastic scattering with a nucleon or nucleus at the EIC or HERA.}
\label{fac0}
\end{figure}

In the DIS process in electron-nucleon/nucleus collisions, an energetic lepton scatters off the nucleon/nucleus target and produces a final state jet. In the correlation measurement, as shown in Fig.~\ref{fac0}, we detect both the lepton and the final state jet, 
\begin{equation}\label{eq:process}
\ell (k)+A(P_A)\to \ell'(k_\ell)+{\rm Jet} (P_J)+X \ ,
\end{equation}
where the incoming lepton and hadron carry momenta $k$ and $P_A$, and the outgoing lepton and jet have momenta $k_\ell$ and $P_J$, respectively. We further define the rapidities of the final state lepton and jet as $y_\ell$ and $y_j$ and the respective transverse momenta $k_{\ell\perp}$ and $P_{J\perp}$. All of these kinematic variables are defined in the center of mass frame of the incoming lepton and hadron. At hadron colliders, dijet correlations have also been  studied~\cite{Abazov:2004hm,Khachatryan:2011zj,daCosta:2011ni,Adamczyk:2013jei,Adamczyk:2017yhe} in terms of similar observables.

Inclusive jet and dijet production have been studied extensively at the HERA collider. However, all these measurements were performed in the center of mass frame of the virtual photon and the nucleon~\cite{Abramowicz:2017ful,Abramowicz:2012jz,Abramowicz:2010ke}. For the proposed lepton-jet correlation in the center of mass frame of the incoming lepton and hadron, the leading order contribution leads to the final state jet and lepton which are back-to-back in the transverse plane, i.e., the azimuthal angular distribution will peak around $\phi=\pi$. The intrinsic transverse momentum of the quark in the nucleon and higher order gluon radiation will induce an imbalance
between these final state particles. In the correlation limit that the imbalance transverse momentum $q_\perp=|\vec{k}_{\ell\perp}+\vec{P}_{J\perp}|$ is much smaller than the lepton (and jet) transverse momentum, we can
factorize the differential cross section in terms of the transverse momentum dependent (TMD) quark distribution~\cite{Collins:1981uk,Collins:1981uw,Collins:1984kg,Ji:2004wu,Collins:2011zzd,GarciaEchevarria:2011rb,Chiu:2012ir} and the soft factor associated with the final state jet. This factorization is similar to that for the semi-inclusive
hadron production in DIS (SIDIS)~\cite{Mulders:1995dh,Boer:1997nt,Bacchetta:2006tn}, where a final state TMD fragmentation will contribute as well. 

More recently, a number of interesting proposals and detailed studies of jet physics at the EIC have emerged~\cite{Gutierrez-Reyes:2018qez,Gutierrez-Reyes:2019vbx,Gutierrez-Reyes:2019msa,Aschenauer:2019uex,Page:2019gbf,Arratia:2019vju,Kang:2020xyq,Arratia:2020ssx,Arratia:2020azl,Arratia:2020nxw,Hinderer:2015hra,DAlesio:2017nrd,Boughezal:2018azh,Borsa:2020ulb}. In particular, it was shown~\cite{Gutierrez-Reyes:2019vbx,Gutierrez-Reyes:2019msa,Aschenauer:2019uex,Page:2019gbf,Arratia:2019vju,Kang:2020xyq,Arratia:2020ssx,Arratia:2020azl,Arratia:2020nxw} that the systematic analysis of jet observables including the lepton-jet correlation can be utilized for the tomography of the nucleon/nucleus. Together with these investigations, our studies in this paper and those in Ref.~\cite{Liu:2018trl} will play an important role to motivate further jet physics research at the EIC.

The rest of this paper is organized as follows. In Sec. II, we will present a detailed study on the TMD factorization for the lepton-jet correlation in $ep$ and $eA$ collisions. We will introduce the soft factor associated with the jet and the TMD quark distribution. One-loop calculations for both the unpolarized and single-transverse-spin dependent differential cross sections will be evaluated explicitly. The factorization will be demonstrated as well. In Sec. III, we perform phenomenological studies for the relevant kinematics at HERA. In particular, the TMD quark distribution at small-$x$ will be investigated by utilizing the currently known parametrization of the TMD quark distributions. We emphasize potentially important constraints of these distributions at small-$x$ from data. This also highlights the impact of future measurements at the EIC. Finally, we summarize our paper in Sec.~IV.

\section{TMD Factorization at Low Imbalance Transverse Momentum}

Including the spin asymmetry, the differential cross sections for the process of Eq.~(\ref{eq:process}) can written as
\begin{eqnarray}
&&\frac{d^5\sigma(\ell p\to \ell' J)}{dy_\ell d^2k_{\ell\perp} d^2q_\perp}=\nonumber\\
&&~~~~\sigma_0\left(W_{UU}(Q;q_\perp)+\epsilon^{\alpha\beta}S_\perp^\alpha W_{UT}^\beta (Q;q_\perp)\right)\,.
\end{eqnarray}
Here the first term corresponds to the spin-averaged cross section, and the second to the transverse single spin dependent contribution, where $\epsilon^{\alpha\beta}$ is defined as $\epsilon^{\alpha\beta\mu\nu}P_{A\mu}k_{\nu}/P_A\cdot k$ with the convention $\epsilon^{0123}=1$. When $q_\perp\ll Q$, where $Q$ is the virtuality of the exchanged photon, the structure functions $W_{UU,UT}$ can be formulated in terms of TMD factorization. In this section, we will demonstrate this factorization by an explicit calculation at one-loop order in the correlation limit, i.e., $q_\perp\ll Q$. We work in the collinear framework, where the incoming quark distribution and quark-gluon-quark correlation function (the Qiu-Sterman matrix element defined below) are the basic ingredients for the unpolarized and single transverse spin dependent differential cross sections, respectively.

At leading order the lepton scatters off the quark through a $t$-channel virtual photon exchange. The virtuality of the photon defines the hard-scattering process. The differential cross section can be written as
\begin{equation}
\frac{d^5\sigma^{(0)}}{dy_\ell d^2 k_{\ell \perp} d^2q_{\perp}}=\sigma_0 xf_q(x) \delta^{(2)}(q_\perp) \ ,
\end{equation}
for the unpolarized case, where the prefactor is given by
\begin{equation}
    \sigma_0=\frac{\alpha_e^2e_q^2}{\hat s Q^2}\frac{2(\hat s^2+\hat u^2)}{Q^4}\, .
\end{equation}
In the above equation, $x$ represents the momentum fraction of the incoming nucleon carried by the quark, $f_q(x)$ for the quark distribution function. The Mandelstam variables $\hat s$, $\hat t$ and $\hat u$ are defined as usual for the partonic sub-process, in particular, we have $\hat t=(k_\ell-k)^2=-Q^2$. At this order, the transverse momenta of the final state lepton and jet are balanced which is indicated by the delta function in the above equation. In addition, the rapidities of the two final state particles are also correlated,
\begin{eqnarray}
&&1=\frac{k_{\ell \perp}}{\sqrt{S_{ep}}}\left(e^{y_\ell}+e^{y_J}\right)\, ,\\
&&x=\frac{k_{\ell\perp}}{\sqrt{S_{ep}}}\left(e^{-y_\ell}+e^{-y_J}\right) \, ,
\end{eqnarray}
where $S_{ep}$ is the center of mass energy squared of the incoming lepton and the nucleon, and $y_J$ is the jet rapidity in the center of mass frame. 

The TMD factorization for the two structure functions $W_{UU,UT}$ can be expressed in terms of the Fourier transform with respect to the transverse momentum $q_\perp$ as
\begin{eqnarray}
W_{UU}(Q;q_\perp)&=&\int\frac{d^2b_\perp}{(2\pi)^2}e^{i\vec{q}_\perp\cdot \vec{b}_\perp} \widetilde{W}_{UU}(Q;b_\perp) \, ,\\
W_{UT}^\alpha(Q;q_\perp)&=&\int\frac{d^2b_\perp}{(2\pi)^2}e^{i\vec{q}_\perp\cdot \vec{b}_\perp} \widetilde{W}_{UT}^\alpha(Q;b_\perp) \, .
\end{eqnarray}
The leading-order Born diagram contributions to $\widetilde{W}_{UU}$ and $\widetilde{W}_{UT}$ are given by
\begin{eqnarray}
\widetilde{W}_{UU}^{(0)}(Q,b_\perp)&=&xf_q(x) \ ,\nonumber\\
\widetilde{W}_{UT}^{(0)\alpha}(Q,b_\perp)&=&\left(\frac{ib^\alpha}{2}\right)xT_F(x,x) \ ,
\end{eqnarray}
where $x$ is defined above, and $f_q(x)$ represents the integrated quark distribution function. The single transverse spin asymmetry comes from the quark Sivers function~\cite{Sivers:1989cc} and the associated twist-three quark-gluon-quark correlation function~\cite{et,qiusterman},
\begin{eqnarray}
T_F(x_1,x_2)& =&
\int\frac{d\xi^-d\eta^-}{4\pi}e^{i(k_{q1}^+\eta^-+k_g^+\xi^-)}
\, \epsilon_\perp^{\beta\alpha}S_{\perp\beta}\nonumber\\
&&\times \left\langle PS|\overline\psi(0){\cal L}(0,\xi^-)\gamma^+
gF_{\alpha}^{\ +}(\xi^-)\right.\nonumber\\
&&\left.\times {\cal L}(\xi^-,\eta^-)
\psi(\eta^-)|PS\right\rangle \ ,
\end{eqnarray}
where 
\begin{align}
x_1&=\,k_{q1}^+/P^+ \,,\\
x_1&=\,k_{q1}^+/P^+ \,,\\
x_g&=\,k_g^+/P^+ = x_2-x_1 \,,
\end{align}
and ${\cal L}$ is the light-cone gauge link making the above definition gauge invariant.

In the following, we are going to derive the one-loop corrections and show that TMD factorization is valid for both the unpolarized and the single transverse spin dependent cross section in the correlation limit. In particular, the collinear divergence will be factorized into the relevant TMD quark distribution, and additional soft gluon radiation into a soft factor associated with the final state jet. The hard factors will be derived at one-loop order based on these results. First, we consider the unpolarized case and second the Sivers asymmetry.

\subsection{The unpolarized Differential Cross Section at One-loop Order}

\begin{figure}[t]
\begin{center}
\includegraphics[width=7cm]{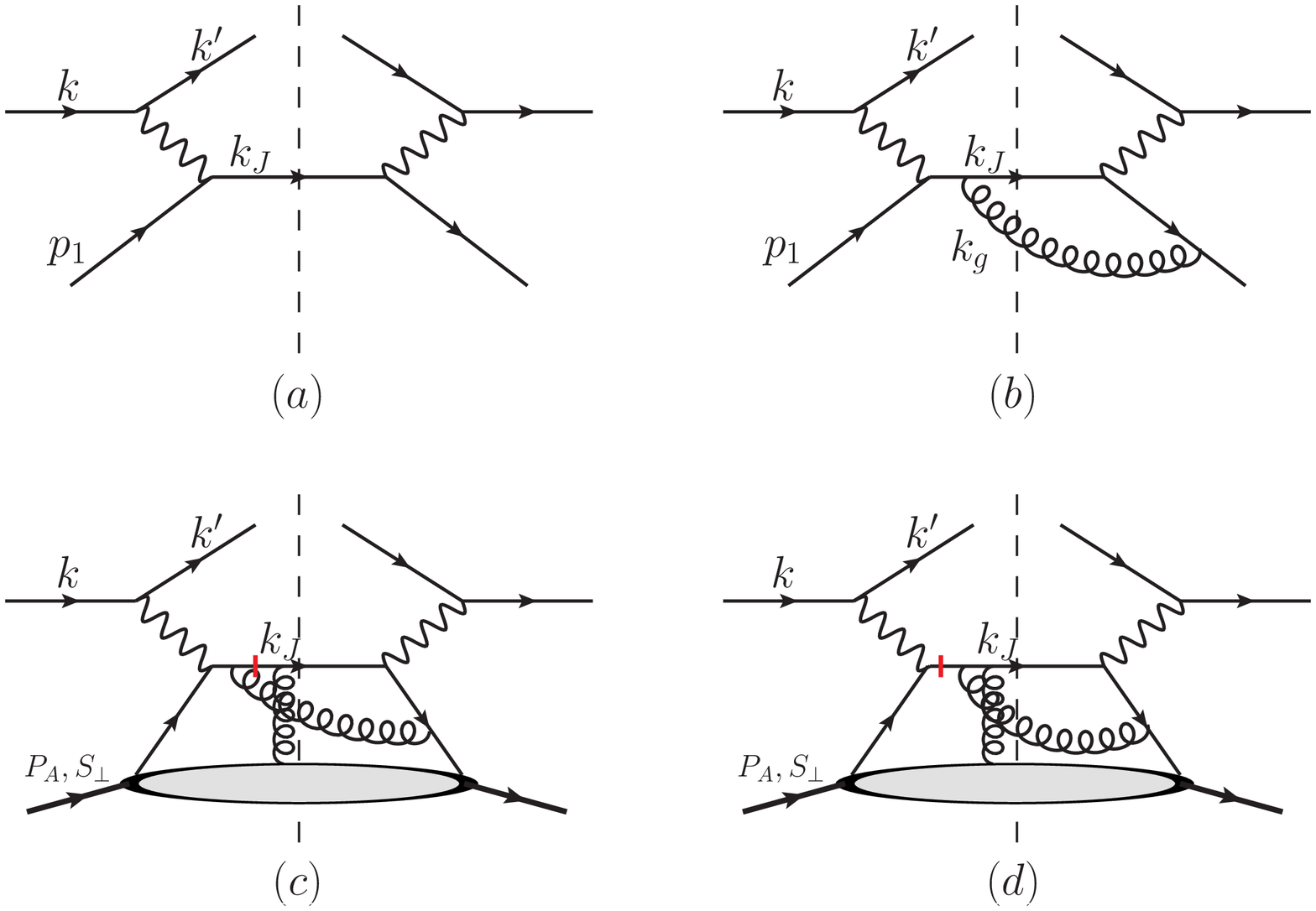}
\end{center}
\caption[*]{Feynman diagrams contributing to the lepton-jet correlation in DIS: (a) the leading order $\ell q\to \ell' q$ scattering; (b) a representative diagram for the soft gluon radiation contribution at one-loop oder; (c) the soft-pole contribution to the single transverse spin asymmetry for this process; (d) the same as (c) but for the hard pole contribution. The lower parts in (c) and (d) represent the twist-three quark-gluon-quark correlation function which depends on the transverse spin of the nucleon. The red vertical lines in these two diagrams indicate the pole contribution from the associated quark propagators. In order to obtain the complete contribution from the hard pole, we have to take into account the diagrams where the vertical gluon is attached to the quark line as shown in (d), the gluon line, and the quark line before the gluon radiation vertex.~\label{fac1}}
\end{figure}

The leading-order diagram lepton-quark scattering diagram is shown in Fig.~\ref{fac1}(a). This leads to a Delta function of $q_\perp$ as $\delta^{(2)}(q_\perp)$, which means that the lepton and quark  in the final state are back-to-back in the transverse plane. At higher orders in perturbative QCD, gluon radiation will contribute a finite transverse momentum. The collinear gluon parallel to the incoming quark is included as part of the TMD quark distribution, whereas those parallel to the final state quark jet are part of jet function. Because the latter does not contribute to a finite $q_\perp$, it will not appear explicitly in the TMD factorization formula. 

The soft gluon radiation is of particular interest, as it shows the structure of the TMD factorization. We show the typical soft gluon radiation diagrams in Fig.~\ref{fac1}(b) for the unpolarized case, and in Fig.~\ref{fac1}(c,d) for transverse spin dependent cross sections. 

The soft gluon radiation is also important to understand how the jet contribution enters the TMD factorization and resummations~\cite{Banfi:2008qs,Mueller:2013wwa,Sun:2014gfa,Sun:2015doa,Sun:2016kkh,Sun:2018icb,Sun:2018beb}. In the process of (1), we have a final state jet, where the soft gluon radiation from the jet will contribute to the imbalance transverse momentum $q_\perp$, from for example the diagram shown in Fig.~\ref{fac1}(b),
\begin{eqnarray}
g^2\int \frac{d^3k_g}{(2\pi)^32E_{k_g}}\delta^{(2)} (q_\perp-k_{g\perp})
C_FS_g(k_J,p_1)\, .
\end{eqnarray}
Here $k_g$ is the momentum of the radiated gluon, and $p_1$ and $k_J$ are the momenta of the initial and final state quark. In addition, $S_g(k_J,p_1)$ is a short-hand notation for 
\begin{equation}
    S_g(k_J,p_1)=\frac{2k_J\cdot p_1}{k_J\cdot k_gp_1\cdot k_g}\, .
\end{equation}
We have to subtract the soft gluon radiation inside the jet cone, which actually belongs to the jet. Therefore, this contribution will depend on the jet size $R$. In previous calculations, the out of cone radiation was derived by assuming a small offshellness for the quark~\cite{Sun:2014gfa,Sun:2015doa,Sun:2016kkh,Sun:2018icb,Sun:2018beb}. Here we apply the subtraction method in dimensional regularization directly. In particular, the jet radius dependent term can be written as
 \begin{eqnarray}
I_R&=&\int\frac{d\xi}{\xi}\frac{d\phi}{2\pi}\frac{\vec{k}_{J\perp}\cdot \vec{k}_{g\perp}}{k_J\cdot k_g} \Theta (\Delta_{k_Jk_g}>R^2)\nonumber\\
&=&\int\frac{d\xi}{\xi}\frac{d\phi}{2\pi}\frac{\vec{k}_{J\perp}\cdot \vec{k}_{g\perp}}{k_J\cdot k_g} \left[1-\Theta (\Delta_{k_Jk_g}<R^2)\right]\, ,
\end{eqnarray}
where $\xi=k_g\cdot p_1/k_J\cdot p_1$, and $\Delta_{k_Jk_g}$ represents
the distance between the two particles with momenta $k_{g}$ and $k_J$. We apply the narrow jet approximation (NJA)~\cite{Mukherjee:2012uz} to derive the $R$ dependence in the above equation. As a result, both terms in the last line can be evaluated in dimensional regularization, and the final result is $I_R=-\frac{1}{\epsilon}\left[1-R^{-2\epsilon}\right]$. We have included a more detailed derivation in Appendix B. Finally, we find the following contribution,
\begin{eqnarray}
S_g(k_J,p_1)&\propto &\frac{\alpha_s}{2\pi^2}\frac{1}{q_\perp^2}\left[\ln\frac{Q^2}{q_\perp^2}+\ln\frac{1}{R^2}+\ln\frac{Q^2}{k_{\ell\perp}^2}\right.\nonumber\\
&&\left.
+\epsilon\left(\frac{1}{2}\ln^2\frac{1}{R^2}\right)\right]\  ,
\end{eqnarray}
where we have kept the $\epsilon$ term for completeness. These terms will contribute a finite term when taking the Fourier transform to $b_\perp$-space. Compared to the results in Refs.~\cite{Sun:2015doa}, we find a finite difference due to the $\epsilon$-term in the above equation. This actually explains the numerical difference found in Refs.~\cite{Sun:2016kkh,Sun:2016kkh,Sun:2018beb} compared to the full NLO calculation. In order to derive the above result, we have averaged over the azimuthal angle of the jet. This average does not factorize and our resummed results given below results will not be accurate beyond next-to-leading logarithmic (NLL) order $\alpha_s^n \ln^n b_\perp$ in the resummed exponent.

In addition, we also have collinear gluon contributions parallel to the incoming quark or the final state quark jet. The former can be expressed as a quark splitting contribution, whereas the latter is formulated with the jet splitting. In this paper, we will follow the narrow jet approximation with the anti-$k_t$~\cite{Cacciari:2008gp} algorithm to compute this contribution. Adding all the above contributions, we obtain the final result in the transverse momentum space,
\begin{eqnarray}
W_{UU}&=&\int\frac{d\xi}{\xi}x'f_q(x')\frac{\alpha_s}{2\pi^2}
C_F\frac{1}{q_\perp^2}\left\{\frac{1+\xi^2}{(1-\xi)_+}+\epsilon(1-\xi)\right.\nonumber\\
&&\left.+\delta(1-\xi)\left[\ln\frac{Q^2}{q_\perp^2}+\ln\frac{1}{R^2}+\ln\frac{Q^2}{k_{\ell\perp}^2}\right.\right.\nonumber\\
&&\left.\left.
+\epsilon\left(\frac{1}{2}\ln^2\frac{1}{R^2}\right)\right]\right\}\, .
\end{eqnarray}
Taking the Fourier transform of the above result to $b_\perp$-space, and adding the contribution from the virtual graph and the jet (see Appendix A), we  find the final result at one-loop order,
\begin{eqnarray}
&&\widetilde{W}_{UU}^{(1)}(Q,b_\perp)=\frac{\alpha_s}{2\pi}C_F\int\frac{d\xi}{\xi}xf_q(x')\left\{\left(-\frac{1}{\epsilon}+\ln\frac{\mu_b^2}{\mu^2}\right)\right.\nonumber\\
&&~~\times {\cal P}_{q\to q}(\xi)+(1-\xi)+\delta(1-\xi)\left[\frac{3}{2}\ln\frac{k_{\ell\perp}^2}{\mu_b^2}\right.\nonumber\\
&&~~\left.\left.-\frac{1}{2}\ln^2\frac{Q^2}{k_{\ell\perp}^2}
-\frac{1}{2}\left(\ln\frac{Q^2}{\mu_b^2}\right)^2-\ln\frac{Q^2}{k_{\ell\perp}^2R^2}\ln\frac{k_{\ell\perp}^2}{\mu_b^2}\right.\right.\nonumber\\
&&~~\left.\left.+\frac{3}{2}\ln\frac{1}{R^2}+3\ln\frac{Q^2}{k_{\ell\perp}^2}-\frac{3}{2}-\frac{2\pi^2}{3}\right]\right\} \ ,\label{euu}
\end{eqnarray}
where $\mu_b=|b_\perp|/c_0$, $c_0=2e^{-\gamma_E}$, and ${\cal P}_{q\to q}(\xi)=\left(\frac{1+\xi^2}{1-\xi}\right)_+$ is the quark splitting kernel. This result can be factorized into the TMD quark distribution 
and the soft factor associated with the final state jet. Most importantly, there are no factorization breaking effects for this process, and the TMD quark distribution is the same as for SIDIS. This in contrast to the dijet production process in hadron-hadron collisions, where TMD factorization is known to be broken at higher orders in perturbation theory~\cite{Boer:2003tx,Qiu:2007ey,Bomhof:2007su,Collins:2007nk,Rogers:2010dm,Bacchetta:2005rm,Vogelsang:2007jk,Catani:2011st,Mitov:2012gt,Schwartz:2017nmr,Schwartz:2018obd,Catani:2013tia,Rothstein:2016bsq}.

Therefore, the above one-loop result should be factorized into the following TMD quark distribution,
\begin{eqnarray}
&&f_q^{\textrm{(unsub.)}}(x,k_\perp)=\frac{1}{2}\int
        \frac{d\xi^-d^2\xi_\perp}{(2\pi)^3}e^{-ix\xi^-P^++i\vec{\xi}_\perp\cdot
        \vec{k}_\perp} \nonumber\\
        &&~~~~\times \left\langle
PS\left|\overline\psi(\xi){\cal L}_{n}^\dagger(\xi)\gamma^+{\cal L}_{n}(0)
        \psi(0)\right|PS\right\rangle\ ,\label{tmdun}
\end{eqnarray}
with the future-pointing gauge link, 
\begin{equation}
{\cal L}_{n}(\xi) \equiv \exp\left(-ig\int^{\infty}_0 d\lambda
\, v\cdot A(\lambda n +\xi)\right)\ .
\end{equation}
This definition contains a light-cone singularity from higher order corrections. The regulation and subtraction defines the scheme of the TMD distributions. Here, as an example, we follow the Collins 2011 scheme which includes a soft factor subtraction as~\cite{Collins:2011zzd},
\begin{eqnarray}
\tilde f_{q}^{\textrm{(sub.)}}(x,b_\perp,\mu_F,\zeta_c)&=&\tilde f_q^{\textrm{(unsub.)}}(x,b_\perp)\nonumber\\
&&\times \sqrt{\frac{\tilde{S}_2^{\bar
n,v}(b_\perp)}{\tilde{S}_2^{n,\bar n}(b_\perp)\tilde{S}_2^{n,v}(b_\perp)}} \, . \label{jcc}
\end{eqnarray}
Here $b_\perp$ is the Fourier conjugate variable with respect to
the transverse momentum $k_\perp$, $\mu_F$ is the factorization
scale and we have $\zeta_c^2=x^2(2v\cdot P)^2/v^2=2(xP^+)^2e^{-2y_n}$ with
$y_n$ the rapidity cutoff in the Collins 2011 scheme. The second factor represents the 
soft factor subtraction with the light-front vectors $n=(1^-,0^+,0_\perp)$,
$\bar n=(0^-,1^+,0_\perp)$, and $v$ is off-light-front $v=(v^-,v^+,0_\perp)$ with $v^-\gg v^+$. 
The light-cone singularity of the un-subtracted
TMDs is cancelled by the soft factor as in Eq.~(\ref{jcc}) with $\tilde{S}^{v_1,v_2}$
defined as
\begin{equation}
\tilde{S}_2^{v_1,v_2}(b_\perp)={\langle 0|{\cal L}_{v_2}^\dagger(b_\perp) {\cal
L}_{v_1}^\dagger(b_\perp){\cal L}_{v_1}(0){\cal
L}_{v_2}(0)  |0\rangle   }\, . \label{softg}
\end{equation}
The one-loop results for the TMD quark distributions can be found for example in Refs.~\cite{Collins:2011zzd,Sun:2013hua}. For convenience, we list their result
\begin{eqnarray}
f_q^{\textrm{(sub.)}}(x,q_\perp)&=&\frac{\alpha_s}{2\pi^2}\frac{C_F}{q_\perp^2}\int\frac{dx'}{x'}f_q(x')\left[\frac{1+\xi^2}{(1-\xi)_+}\right.\nonumber\\
&&\left. +\epsilon(1-\xi)+\delta(1-\xi)\ln\frac{\zeta_c^2}{q_\perp^2}\right] \, .
\end{eqnarray}
In Fourier transform $b_\perp$-space, we have
\begin{eqnarray}
&&\tilde f_q^{\textrm{(sub.)}}(x,b_\perp)=\frac{\alpha_s}{2\pi}C_F\int\frac{d\xi}{\xi}f_q(x')\left\{\left(-\frac{1}{\epsilon}+\ln\frac{\mu_b^2}{\bar\mu^2}\right)\right.\nonumber\\
&&~~~\times {\cal P}_{q\to q}(\xi)+(1-\xi)+\delta(1-\xi)\left[\frac{3}{2}\ln\frac{\mu^2}{\mu_b^2}\right.\nonumber\\
&&~~~\left.\left.+\frac{1}{2}\left(\ln\frac{\zeta_c^2}{\mu^2}\right)^2
-\frac{1}{2}\left(\ln\frac{\zeta_c^2}{\mu_b^2}\right)^2\right]\right\} \ .
\end{eqnarray}
The soft factor associated with the final state jet can be defined accordingly
\begin{eqnarray}
\tilde{S}_J(b_\perp,\mu_F)=\frac{\tilde{S}_{n_1,\bar n}(b_\perp)}{\sqrt{\tilde{S}_{n,\bar n}(b_\perp)}}\ ,\label{softfactor}
\end{eqnarray}
where $n_1$ represents the jet direction. A one-loop calculation leads to the following result
\begin{eqnarray}
S_J^{(1)}(k_\perp)=\frac{\alpha_s}{2\pi^2}\frac{1}{q_\perp^2}C_F\left[\ln\frac{\hat t}{\hat u}+\ln\frac{1}{R^2}\right] \ ,
\end{eqnarray}
in the transverse momentum space. In $b_\perp$-space, we obtain, to first order,
\begin{eqnarray}
\tilde{S}_J^{(1)}(b_\perp)=\frac{\alpha_s}{2\pi}\left[-\ln\frac{\hat t}{\hat u R^2}\ln\frac{\mu^2}{\mu_b^2}+\frac{1}{2}\ln^2\frac{1}{R^2}\right]\ .
\end{eqnarray}
From this result, we derive the anomalous dimension at one-loop order,
\begin{eqnarray}
\gamma_s^{(1)}=-C_F\frac{\alpha_s}{2\pi}\ln\frac{\hat t}{\hat u R^2} \ .
\end{eqnarray}
With the above results, we can verify the TMD factorization at this order,
\begin{eqnarray}
\widetilde{W}_{UU}(x,b_\perp)&=&
\tilde f_{q}^{\textrm{(sub.)}}(x,b_\perp,\mu_F,\zeta_c) \tilde{S}_J(b_\perp,\mu_F)\nonumber\\
&&\times H_{\rm TMD}(Q,\mu_F) \ ,\label{facb}
\end{eqnarray}
with the hard factor 
\begin{eqnarray}
H_{\rm TMD}^{(1)}&=&\frac{\alpha_sC_F}{2\pi}\left[-\ln^2\frac{Q^2}{k_{\ell\perp}^2}+\frac{1}{2}\ln^2\frac{1}{R^2}+\frac{3}{2}\ln\frac{1}{R^2}\right.\nonumber\\
&&\left.+3\ln\frac{Q^2}{k_{\ell\perp}^2}-\frac{3}{2}-\frac{2\pi^2}{3}
\right] \ ,
\end{eqnarray}
where we have chosen $\zeta_c^2=\hat s$ and $\mu_F^2=k_{\ell\perp}^2$ to simplify the
expression.

\subsection{The Sivers Asymmetry at One-loop Order}

The above factorization applies to the single transverse spin asymmetry
in the process of (1). In Ref.~\cite{Qiu:2007ey}, it was shown that the collinear
gluon radiation from the polarized nucleon can be factorized into the TMD quark
Sivers function. To establish the complete factorization formalism, we need to 
demonstrate that the soft gluon radiation can be factorized and expressed 
as Eq.~(\ref{softfactor}) as well. The soft gluon radiation comes from the Feynman diagrams 
shown in Fig.~\ref{fac1}. Similar diagrams have been calculated for the SIDIS
process, where the final state fragmentation function will contribute~\cite{Ji:2006br,Koike:2007dg}. In the current case, it is the final state jet contribution. 

To obtain a non-zero single spin asymmetry, we have to take into account final state interaction effects, which generate a phase through a pole from the interference diagrams, as shown for example in Fig.~\ref{fac1}(c) and(d). We have marked the pole places in these diagrams, where \ref{fac1}(c) refers to the so-called soft-pole and \ref{fac1}(d) to  the hard-pole, similar to the SIDIS process which was calculated in Ref.~\cite{Ji:2006br,Koike:2007dg}. There are also soft-fermion pole contributions from $T_F$ and those from the $\widetilde{G}_F$ twist-three function~\cite{Koike:2007dg}. These contributions can be analyzed in a similar manner. In the following, we only consider the soft-gluon and hard-gluon pole contributions. The soft pole corresponds to the case where the vertical gluon carries zero longitudinal momentum fraction of the incoming nucleon when taking the pole, whereas the hard pole corresponds to a non-zero momentum fraction for the gluon. The calculations of these diagrams will be the same as those in Ref.~\cite{Ji:2006br}. Again, in the TMD limit, i.e., the transverse momentum of the radiated gluon $q_\perp$ is much smaller than the hard momentum scale ($k_{\ell \perp}$ in our case), there exist cancellations between the soft- and hard-pole contributions. This is particularly the case for the soft gluon radiation diagrams in Fig.~\ref{fac1}(c) and (d).

For example, the soft-pole diagram of Fig.~\ref{fac1}(c) has a color factor of $-1/2N_c$, and part of the hard-pole contribution (such as Fig.~\ref{fac1}(d)) has a color factor of $1/2N_c$ as well. If we decompose the hard pole contribution into the color factors $C_F$ and $1/2N_c$, we find that the term proportional $1/2N_c$ cancels completely against that from the soft-pole contribution. The final result will only depend on the color factor $C_F$, which is proportional to the following structure,
\begin{eqnarray}
&&g^2C_F\epsilon^{\alpha\beta}S_{\perp\alpha}\int \frac{d^3k_g}{(2\pi)^32E_{k_g}}\delta^{(2)} (q_\perp-k_{g\perp})
\left[S_g(k_J,p_1)\right]^2\nonumber\\
&&~~~\times \left(k_{g\perp\beta}-\xi k_{J\perp\beta}\right)\, .
\end{eqnarray}
Here $S_\perp$ represents the traverse spin vector of the incoming nucleon. The above result contains a collinear divergence associated with the final state quark jet when $k_{g\perp}\sim \xi k_{J\perp}$. As we have shown above, $S_g(k_J,p_1)$ has a single power collinear divergence. Therefore, the spin dependent cross section contribution depends on the combination of $k_{g\perp}-\xi k_{J\perp}$, which will cancel one power of the collinear divergence from $\left[S_g(k_J,p_1)\right]^2$. As a result, we have only one collinear divergence associated with soft gluon radiation from the jet. After integrating over $\xi$, we obtain the following contribution,
\begin{eqnarray}
&&\frac{\alpha_s}{2\pi^2}\frac{\epsilon^{\alpha\beta}S_{\perp\alpha}q_{\perp\beta}}{(q_\perp^2)^2}
\left[\ln\frac{Q^2}{q_\perp^2}+\ln\frac{1}{R^2}+\ln\frac{Q^2}{k_{\ell\perp}^2}\right.\nonumber\\
&&\left.~~~+\epsilon\left(\frac{1}{2}\ln^2\frac{1}{R^2}\right)\right]\,  .
\end{eqnarray}
Adding the collinear gluon radiation which is parallel to the incoming proton, and the contribution from the virtual graph and the jet as for the unpolarized case above, we have the following one-loop result in $b_\perp$-space,
\begin{eqnarray}
&&\widetilde{W}_{UT}^\alpha(Q,b)=\frac{\alpha_s}{2\pi}\frac{ib^\alpha}{2}x\int\frac{d\xi}{\xi}\left\{\left(-\frac{1}{\epsilon}+\ln\frac{\mu_b^2}{\mu^2}\right)
\right.\nonumber\\
&&~~~\times {\cal P}_{qg\to qg}^T\otimes T_F(x,x)-T_F(x',x')\frac{1}{2N_c}(1-\xi)\nonumber\\
&&~~~\left.+\delta(1-\xi)T_F(x,x)C_F\left[-\frac{1}{2}\ln^2\frac{Q^2}{k_{\ell\perp}^2}
-\frac{1}{2}\left(\ln\frac{Q^2}{\mu_b^2}\right)^2\right.\right.\nonumber\\
&&~~~-\ln\frac{Q^2}{k_{\ell\perp}^2R^2}\ln\frac{k_{\ell\perp}^2}{\mu_b^2}+\frac{3}{2}\ln\frac{k_{\ell\perp}^2b^2}{c_0^2}+\frac{3}{2}\ln\frac{1}{R^2}\nonumber\\
&&~~~\left.\left.+3\ln\frac{Q^2}{k_{\ell\perp}^2}-\frac{3}{2}-\frac{2\pi^2}{3}\right]\right\} \, ,\label{eut}
\end{eqnarray}
where the splitting kernel for the Sivers function is obtained from the Qiu-Sterman matrix
element~\cite{Braun:2009mi,Kang:2008ey,Vogelsang:2009pj,Zhou:2008mz,Schafer:2012ra,Scimemi:2019gge,Scimemi:2019gge}.

Again, this can be factorized into the TMD quark Sivers function and the soft factor associated with the jet,
\begin{eqnarray}
\widetilde{W}_{UT}^{\alpha}(x,b_\perp)&=&
\tilde f_{1T}^{\perp\alpha{\textrm{(sub.)}}}(x,b_\perp,\mu_F,\zeta_c) \tilde{S}_J(b_\perp,\mu_F)\nonumber\\
&&\times H_{\rm TMD}(\mu_F) \ ,
\end{eqnarray}
where the soft factor and the hard factor are the same as those in the unpolarized case. Similar to the above case, the quark Sivers function needs a subtraction as well,
\begin{eqnarray}
&&\tilde f_{1T}^{\perp\alpha{\textrm{(sub.)}}}(x,b_\perp,\mu_F,\zeta_c)=\tilde f_{1T}^{\perp\alpha{\textrm{(unsub.)}}}(x,b_\perp)\nonumber\\
&&~~~~~~~~~~~~~~~~~\times \sqrt{\frac{\tilde{S}_2^{\bar
n,v}(b_\perp)}{\tilde{S}_2^{n,\bar n}(b_\perp)\tilde{S}_2^{n,v}(b_\perp)}} \ , \label{jccsivers}
\end{eqnarray}
with the same soft factor subtraction in the Collins 2011 scheme. The explicit one-loop expression can also be found in Ref.~\cite{Sun:2013hua}.  

\subsection{Resummation}

There are large logarithms in the TMD quark distributions and the soft factor associated  with the jet, which can be resummed by solving the relevant evolution equations. The TMD quark distribution has been studied extensively in recent years and  can be applied in our case. With resummation effects taken into account, the final result can be written as
\begin{eqnarray}
\widetilde{W}_{UU}^{(res.)}&=&x\tilde{f}_{q}(x,b_\perp,\mu_F=k_{\ell\perp},\zeta_c=\sqrt{\hat s})e^{-\Gamma_s}{\cal C}_{UU}\ ,\nonumber\\
\widetilde{W}_{UT}^{(res.)\alpha}&=&x\tilde{f}_{1T}^{\perp\alpha}(x,b_\perp,\mu_F=k_{\ell\perp},\zeta_c=\sqrt{\hat s})e^{-\Gamma_s}{\cal C}_{UT}\, ,\label{vq} \nonumber \\
\end{eqnarray}
where $f_{q}$ and $f_{1T}^\perp$ represent the standard unpolarized quark TMD
distribution and the Sivers function, respectively. Here, $\Gamma_s$ is associated with the soft factor due to the jet and it is given by: 
\begin{equation}
\Gamma_s=\int_{c_0^2/b_\perp^2}^{k_\perp^2}
\frac{d\mu^2}{\mu^2}\gamma_s \, ,
\end{equation}
where $\gamma_s^{(1)}=\alpha_sC_F\ln(\hat t/\hat u R^2)/2\pi$. The one-loop expressions for ${\cal C}_{UU}$ and ${\cal C}_{UT}$ are given by
\begin{eqnarray}
{\cal C}_{UT}&=&{\cal C}_{UU}=\frac{\alpha_s}{2\pi}C_F\left[-\ln^2\frac{Q^2}{k_{\ell\perp}^2}+3\ln\frac{Q^2}{k_{\ell\perp}^2}\right.\nonumber\\
&&\left.+\frac{3}{2}\ln\frac{1}{R^2}-\frac{3}{2}-\frac{2\pi^2}{3}\right] \ .
\end{eqnarray}

\subsection{Non-global Logarithms}

For the lepton-jet correlation in DIS, non-global logarithms (NGLs)~\cite{Dasgupta:2001sh,Dasgupta:2002bw} have to be taken into account. They start contributing at the order ${\cal O}(\alpha_s^2)$. See Appendix C for a detailed derivation. (A similar calculation was performed in Ref.~\cite{Banfi:2003jj}). See also Ref.~\cite{Chien:2019gyf}. The first non-zero contribution is given by
\begin{equation}
    \tilde{{\cal S}}_{\mathrm{\rm NGL}}^{(2)}(b_\perp)=-C_F\frac{C_A}{2}\left(\frac{\alpha_s}{\pi}\right)^2\frac{\pi^2}{24}\ln^2 \left(\frac{k_{\ell\perp}^2b_\perp^2}{c_0^2}\right) \, .
\end{equation}
The resummation of these NGLs is more complicated than that of the global logarithms in the resummation formula in Eq.~(\ref{vq}). For the kinematics we are interested in the NGL contribution is very small and will not be included in the numerical studies presented below. We leave more detailed phenomenological studies for future work.

\section{Phenomenological Studies and Comparison to HERA Data}

Phenomenological results for the lepton-jet correlation at the EIC have been shown in Ref.~\cite{Liu:2018trl}. In particular, the single transverse spin asymmetries for this process have been shown to directly probe the quark Sivers function, whereas the measurement of $P_T$-broadening effects in $eA$ collisions will be a great opportunity to explore cold nuclear matter effects through hard probes.

In this section, we will take the opportunity that the existing experimental data from the HERA collider has been re-analyzed to study the lepton-jet correlation~\cite{Amilkar}. We will compare to these preliminary analyses and comment on the implications of the experimental measurement. This will serve as an important cross check of our formalism and may also indicate constraints on the small-$x$ modification of the TMD quark distribution in the proton. This shall provide an important guideline for future measurements at the EIC.

The TMD quark distribution takes the form~\cite{Prokudin:2015ysa}, 
\begin{eqnarray}
\widetilde{f}_{q}(x,b_\perp,\mu_F=k_{\ell\perp},\zeta_c=\sqrt{\hat s})= {\mathrm{e}}^{-{ {S}_{\rm pert}^q(b_*)}-{S}_{\rm NP}^q(b_\perp)}\nonumber\\
\times\sum_iC_{q/i}(x,\mu_b/\mu)\otimes f_i(x,\mu_b) , \label{tmdqfc}
\end{eqnarray}
where $b_*=b_\perp/\sqrt{1+b_\perp^2/b_{\rm max}^2}$ with $b_{\rm max}=1.5$~GeV$^{-1}$, and $f_i(x,\mu)$ is the integrated parton distribution. The Sudakov form factor for the quark is given by
\begin{equation}
S_{\rm pert}^q(b_\perp)=\int_{\mu_b^2}^{k_{\ell\perp}^2}
\frac{d\mu^2}{\mu^2}\left[A_q\big(\alpha_s(\mu)\big)\ln\frac{\hat s}{\mu^2}+B_q\big(\alpha_s(\mu)\big)\right] \ ,
\end{equation}
with $A_q=\frac{\alpha_s}{2\pi}C_F$, $B_q^{(1)}=-\frac{\alpha_s}{\pi}\,\frac{3}{2}C_F$, and where for simplicity we take the leading order expression for the coefficient function $C$. We use the non-perturbative parametrization of Refs.~\cite{Su:2014wpa,Prokudin:2015ysa}:
\begin{equation}
    S_{\rm NP}^q=0.106\, b_\perp^2+0.42\ln(Q/Q_0)\ln(b_\perp/b_*) \ , \label{np}
\end{equation}
with $Q_0^2=2.4$~GeV$^2$.

For the HERA measurement of Ref.~\cite{Amilkar}, the kinematics are as follows: $Q^2>10\;{\rm GeV}^2$; $30>P_{J\perp}>2.5\rm$~GeV, and the pseudo-rapidity of the jet is $|\eta_j|<1$ in the Lab frame. From these kinematics, it was found that the lepton-jet production is dominated by small-$x$ around $10^{-3}$. Before we compare to the experimental data, it is interesting to study the behavior of the TMD quark distribution using the known parametrizations. In Ref.~\cite{Su:2014wpa,Prokudin:2015ysa}, the TMD formalism was applied to describe the existing Drell-Yan type processes, and no $x$-dependence of the non-perturbative form factor was found by means of a global analysis. 

\begin{figure}[t]
\begin{center}
\includegraphics[width=7cm]{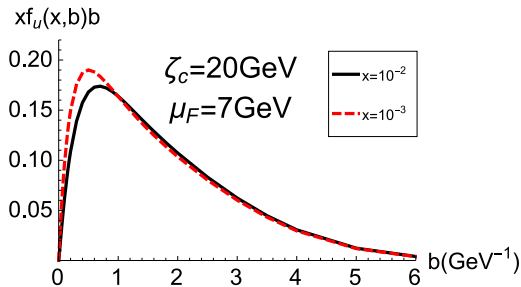}
\end{center}
\caption[*]{The up quark TMD distribution as function of the Fourier transform variable $b_\perp$ for different $x$. We have fixed the factorization scale $\mu_F=7~\rm GeV$ and energy scale $\zeta_c=20~\rm GeV$, which correspond to the kinematics of the HERA data~\cite{Amilkar}. }
\label{tmdx}
\end{figure}

In Fig.~\ref{tmdx}, we show, as an example, the up quark TMD as a function of $b_\perp$ for two different values of $x$. Taking similar kinematics as relevant for the HERA data~\cite{Amilkar}, we fix the factorization scale $\mu_F=7~\rm GeV$ and the energy scale $\zeta_c=20~\rm GeV$. From this figure, we can see that the TMD quark distribution does evolve with $x$. However, the evolution is mild, because it solely comes from the collinear scale dependence of the TMD quark distribution in Eq.~(\ref{tmdqfc}). This, of course, is also because the Drell-Yan type of data in the global analysis of Refs.~\cite{Su:2014wpa,Prokudin:2015ysa} found no $x$-dependence of the non-perturbative form factor. 

\begin{figure}[t]
\begin{center}
\includegraphics[width=7cm]{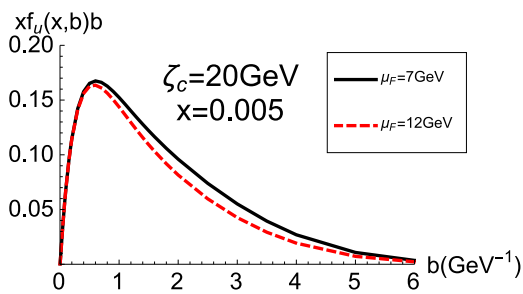}
\end{center}
\caption[*]{The up quark TMD distribution as a function of the Fourier transform variable $b_\perp$ for different values of the factorization scale $\mu_F$. We have fixed $x=0.005$ and the energy scale $\zeta_c=20~\rm GeV$.}
\label{tmdxkt}
\end{figure}

In Fig.~\ref{tmdxkt}, we show the dependence of the TMD quark distribution on the factorization scale $\mu_F$ at small-$x$ which is relevant for the HERA data. It is interesting to note that in the range of the transverse momentum of the jet, the scale dependence is not very strong.

\begin{figure}[t]
\begin{center}
\includegraphics[width=7cm]{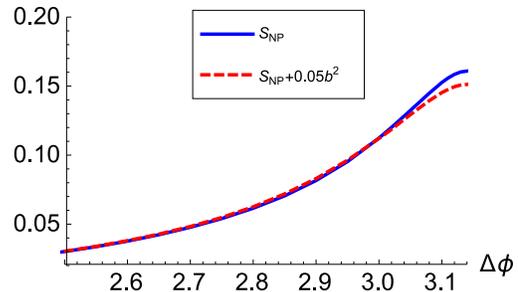}
\end{center}
\caption[*]{The azimuthal angular correlation between the final state lepton and jet for the HERA kinematics. For the numerical results, we have chosen an average jet transverse momentum of $5~\rm GeV$ at mid-rapidity in the Lab frame to evaluate the TMD quark distributions. The blue curve represents the result from the default parametrization of the non-perturbative form factor of Ref.~\cite{Su:2014wpa,Prokudin:2015ysa} as in Eq.~(\ref{np}). For the red-dashed curve, we include an additional small-$x$ contribution as shown in Eq.~(\ref{smallx0}).}
\label{angular}
\end{figure}

To compare to the HERA measurement, we take an average transverse momentum for the jet $P_{J\perp}=5~\rm GeV$ with rapidity $\eta_j=0$ to evaluate the TMD quark distributions. With that, we plot the azimuthal angular distribution between the final state lepton and the jet in Fig.~\ref{angular}. The blue curve represents the prediction with the TMD quark distribution in Eq.~(\ref{tmdqfc}) with the non-perturbative form factor of Eq.~(\ref{np}). In the calculations, we neglect the non-perturbative contribution from the soft factor associated with the jet, i.e., the $S_J$ factor in the factorization formula of Eq.~(\ref{facb}). We expect this part to be smaller than that from the TMD quark distributions. Future experiments can help to constrain this contribution, especially through the jet radius dependence of the correlation measurement.

We further notice that within the HERA kinematics, the azimuthal angular correlation could be sensitive to gluon saturation effects~\cite{McLerran:1993ni,McLerran:1994vd}. Gluon saturation will modify the TMD quark distribution, see, for example, a calculation in Ref.~\cite{Marquet:2009ca}. To illustrate how this affects the azimuthal angular correlation between the final state lepton and the jet, we include a small-$x$ modification of the non-perturbative part
\begin{equation}
    S_{\rm NP}^q\rightarrow S_{\rm NP}^q+\langle\delta q_\perp^2\rangle b^2/4 \ , \label{smallx0}
\end{equation}
with $\langle \delta q_\perp^2\rangle=0.2~\rm GeV^2$ as an example. This modification is qualitatively consistent with the observation in Ref.~\cite{Marquet:2009ca} that transverse momentum broadening arises at small-$x$ due to gluon saturation effects. We plot this prediction (red curve) in Fig.~\ref{angular}. We conclude that small-$x$ effects can lead to a sizable numerical impact. We hope that future measurements can help to test these predictions and impose constraints also on the small-$x$ contribution to the TMDs.

\section{Conclusions}

In this paper, we have presented a detailed derivation of the TMD factorization for the azimuthal angular correlation between the final state lepton and jet in DIS processes. An explicit one-loop calculation was carried out, and the factorization has been verified accordingly. We have also derived the single-transverse spin asymmetry for this process, which depends on the quark Sivers function. The calculations are performed within the twist-three framework and the Sivers function corresponds to the Qiu-Sterman matrix element. Based on the factorization formula, we further derived an all order resummation. 

The factorization formalism demonstrates the striking simplicity of this process, where the total transverse momentum of the final state lepton and jet depends on the TMD quark distribution plus a soft factor associated with the jet. This will provide an important channel to investigate the TMD quark distribution in $ep$ and $eA$ collisions. Some recent phenomenological studies of this correlation have shown very promising results for the future EIC~\cite{Arratia:2019vju,Arratia:2020azl}. We expect more research along this direction in the near future. 

We have also carried out a phenomenological study on the TMD quark distribution for HERA kinematics~\cite{Amilkar}. The kinematics at HERA are sensitive to the small-$x$ region, where the TMD parton distributions are not well constrained. Therefore, the comparison between theory and experiment in this region could provide a potential signal for the $x$-dependence of the TMD quark distributions. We look forward to comparing our theory predictions to the experimental data, which are soon to be published.

\vspace{2em}

{\it Acknowledgment.}  We thank Amilkar Quintero and Bernd Surrow for communications concerning the HERA data. This work is partially supported by the U.S. Department of Energy, Office of Science, Office of Nuclear Physics, under contract number DE-AC02-05CH11231, and by the U.S. National Science Foundation under Grant No. PHY-1417326. This study was supported by Deutsche Forschungsgemeinschaft (DFG) through the Research Unit FOR 2926 (project number 40824754).

\appendix

\section{Soft Gluon Radiation Associated with the Jet}

Following previous derivations~\cite{Sun:2015doa}, we can calculate the soft gluon radiation associated with the jet as
\begin{eqnarray}
&&\int \frac{d^3k_g}{2E_{k_g}}\delta^{(2)} (q_\perp-k_{g\perp})S_g(k_J,p_1) =\frac{\alpha_s}{2\pi^2}\frac{1}{q_\perp^2}\left[\ln\frac{\hat s}{q_\perp^2}\right.\nonumber\\
&&~~\left.+\ln\frac{1}{R_1^2}+\ln\left(\frac{\hat t}{\hat u}\right)
+\epsilon\left(\frac{1}{2}\ln^2\frac{1}{R_1^2}+\frac{\pi^2}{6}\right)\right]\ ,\label{oldsoft}
\end{eqnarray}
where $\hat s$, $\hat t$ and $\hat u$ are define in Sec.~II. To arrive at the above result, an approximation was made in Ref.~\cite{Sun:2015doa}, where an off-shellness was imposed for the jet momentum: $k_J^2\sim k_{J\perp}^2R^2$ with the jet size $R$. In the following, we apply the subtraction method to derive the above result without any approximation. 

First, we notice that the contribution due to soft gluon radiation can be written as
\begin{eqnarray}
S_g(k_J,p_1)+S_g(k_J,p_2)&=&\frac{4}{k_{g\perp}^2}+\frac{4}{k_{g\perp}^2}\frac{\vec{k}_{J\perp}\cdot \vec{k}_{g\perp}}{k_J\cdot k_g} \, ,\;\;\;
\end{eqnarray}
where the first term contributes to the double logarithms and
corresponds to the
first term in the bracket of Eq.~(\ref{oldsoft}). To calculate the second term, we define 
\begin{equation}
I(R)=\int\frac{d\xi}{\xi}\frac{d\phi}{2\pi}\frac{\vec{k}_{J\perp}\cdot \vec{k}_{g\perp}}{k_J\cdot k_g} \, ,
\end{equation}
where $\xi$ is the longitudinal momentum of $k_g$ with respect to $k_J$, $\xi=k_g\cdot p_1/k_J\cdot p_1$ and $\phi$ is the azimuthal angle between $k_{g\perp}$ and $k_{J\perp}$. As mentioned above, we will further average over this angle to obtain the final result. In the NJA, $I(R)\to \ln(1/R^2)$. 

Similarly, we find that
\begin{eqnarray}
S_g(k_J,p_1)-S_g(k_J,p_2)
&=&\frac{4}{k_{g\perp}^2}\frac{k_J^+k_g^--k_J^-k_g^+}{k_J\cdot k_g}  \, .
\end{eqnarray}
It is interesting to notice that the above term does not contain a divergence associated with the jet. Therefore, we can integrate over the longitudinal momentum of the gluon
\begin{equation}
\int\frac{d\xi}{\xi}\frac{d\phi}{2\pi}\frac{k_J^+k_g^--k_J^-k_g^+}{k_J\cdot k_g} =\ln\frac{\hat t}{\hat u}\, .
\end{equation}
Therefore, the only contribution to the soft gluon radiation where the jet radius
and the jet algorithm appear are given by the integral $I(R)$. This integral can be related
to the jet contribution at one-loop order, which also depends on the
jet algorithm. Previously, as discussed in the above section, an approximation
was made to carry out the integral. 
In the following we derive the result for $I(R)$ with the subtraction method
\begin{eqnarray}
I(R)&=&\int\frac{d\xi}{\xi}\frac{d\phi}{2\pi}\frac{\vec{k}_{1\perp}\cdot \vec{k}_{g\perp}}{k_1\cdot k_g} \Theta (\Delta_{k_1k_g}>R^2)\nonumber\\
&=&\int\frac{d\xi}{\xi}\frac{d\phi}{2\pi}\frac{\vec{k}_{1\perp}\cdot \vec{k}_{g\perp}}{k_1\cdot k_g} \left[1-\Theta (\Delta_{k_1k_g}<R^2)\right]\, ,
\end{eqnarray}
where the first term is similar to the global-soft, and the second term is similar to the collinear-soft contribution. 

First, we notice that
\begin{equation}
k_1\cdot k_g=k_{1\perp}k_{g\perp}\left[\cosh(\Delta Y)-\cos (\Delta\phi)\right] \ ,
\end{equation}
where $\Delta Y$ and $\Delta\phi$ are rapidity and angular separation between $k_g$ and $k_J$. Substituting the above into $I(R)$, we find that
\begin{eqnarray}
I(R)&=&\frac{1}{2\pi}\int d\Delta Yd\Delta\phi \frac{\cos(\Delta\phi)}{\cosh(\Delta Y)-\cos (\Delta\phi)}\nonumber\\
&&~~\times 
\left[1-\Theta (\Delta_{k_1k_g}<R^2)\right]\nonumber\\
&=&I_G-I_{cs}(R)\ .
\end{eqnarray}
The global-soft term can be calculated as
\begin{eqnarray}
I_G&=&\frac{1}{2\pi}\int d\Delta Yd\Delta\phi \frac{\cos(\Delta\phi)}{\cosh(\Delta Y)-\cos (\Delta\phi)}\nonumber\\
&=&\frac{1}{\int_0^\pi d\phi \sin^{-2\epsilon}(\phi)}\int_0^\pi d\phi \sin^{-2\epsilon}(\phi)\nonumber\\
&&~~\times \int_0^\infty dy\frac{\cos(\phi)}{\cosh(y)-\cos(\phi)}\nonumber \\
&= &-\frac{1}{\epsilon} \, .
\end{eqnarray}
For the collinear-soft term, we can apply the narrow jet approximation $R\ll 1$, 
\begin{eqnarray}
I_{cs}(R)&=&\frac{1}{2\pi}\int d\Delta Yd\Delta\phi \frac{\cos(\Delta\phi)}{\cosh(\Delta Y)-\cos (\Delta\phi)}\nonumber\\
&&~~\times \Theta (\Delta_{k_1k_g}<R^2)\nonumber\\
&=&\frac{R^{-2\epsilon}}{\int_0^\pi d\phi \sin^{-2\epsilon}(\phi)}\int_0^1 d\phi (\phi)^{-2\epsilon}\nonumber\\
&&~~\times \int_0^{\sqrt{1-\phi^2}} dy\frac{1}{y^2+\phi^2} \nonumber\\
&=&R^{-2\epsilon}\left(-\frac{1}{\epsilon}\right)\, .
\end{eqnarray}
Therefore, $I(R)$ has a rather simple structure
\begin{equation}
I(R)=-\frac{1}{\epsilon}\left[1-R^{-2\epsilon}\right]=\ln\frac{1}{R^2}+\epsilon\frac{1}{2}\ln^2\frac{1}{R^2} \, .
\end{equation}
Certainly, the leading pole cancels and we are left with a term proportional $\ln(1/R^2)$. 

Substituting the above result into the original calculation, we find that there should be no $\pi^2/6$ 
term in the terms $\sim \epsilon$ in Eq.~(\ref{oldsoft}). This applies to all the soft gluon radiation contributions associated with the jet calculated in Refs.~\cite{Sun:2015doa}, and solves the puzzle found in previous calculations of Refs.~\cite{Sun:2016kkh,Sun:2016kkh,Sun:2018beb}.

\section{Virtual contribution and the jet contribution}

The virtual graph has the following contribution in the $\overline{\rm MS}$ scheme,
\begin{equation}
\Gamma^v=\frac{\alpha_s}{2\pi}C_F\left(\frac{\mu^2}{Q^2}\right)^\epsilon \left\{-\frac{2}{\epsilon^2}-\frac{3}{\epsilon}-8\right\} \, .
\end{equation}
The jet contribution is is given by,
\begin{eqnarray}
J_q&=&\frac{\alpha_s}{2\pi}C_F\left(\frac{\mu^2}{P_{J\perp}^2}\right)^\epsilon\left\{
\frac{1}{\epsilon^2}+\frac{1}{\epsilon}\ln\frac{1}{R^2}+\frac{3}{2}\frac{1}{\epsilon}\right.\nonumber\\
&&\left.+\frac{1}{2}\ln^2\frac{1}{R^2}
+\frac{3}{2}\ln\frac{1}{R^2}+I_q'\right\}\, ,
\end{eqnarray}
where $I_q'$ for anti-k$_T$ jets is defined as
\begin{equation}
I_q'=\frac{13}{2}-\frac{2}{3}\pi^2 \, .
\end{equation}

\section{Non-global Logarithms}

NGLs start to contribute at two-loops~\cite{Dasgupta:2001sh,{Dasgupta:2002bw},{Banfi:2003jj}}. These contributions arise from the configuration where one of the soft gluons inside the jet while the second gluon is outside of jet. For the differential cross section that we are studying in this paper, these soft gluon emissions will contribute to a finite transverse momentum. In this section, we will discuss the emission of two soft gluons that leads to the NGLs which need to be included in addition to the above resummation formalism.

The two soft gluon emission matrix element squared can be written as~\cite{Dasgupta:2002bw}
\begin{eqnarray}
W_2&=&C_F^2S_g(p_1,k_J;k_1)S_g(p_1,k_J;k_2)\nonumber\\
&&+S_g(p_1,k_J;k_1)\frac{C_FC_A}{2}\left[S_g(p_1,k_1;k_2)\right.\nonumber\\
&&\left.+S_g(k_J,k_1;k_2)-S_g(p_1,k_J;k_2)\right] \, , \label{soft1}
\end{eqnarray}
for strong ordering of $k_1\gg k_2$, where $k_1$ and $k_2$ are momenta for the two radiated gluons, and $S_g(p_1,p_2;k)$ is defined as
\begin{equation}
S_g(p_1,p_2;k)=\frac{2p_1\cdot p_2}{p_1\cdot k p_2\cdot k} \, .
\end{equation}
The NGL contribution is obtained from the second term in Eq.~(\ref{soft1}) with color factor $C_FC_A$. It can be derived by integrating over the phase space of $k_1$ and $k_2$ with a delta function constraining the transverse momentum,
\begin{eqnarray}
{\cal S}_{\rm NGL}^{(2)}&=&\frac{C_FC_Ag^4}{2}\int\frac{d^3k_1}{(2\pi)^32E_{k_1}}
\frac{d^3k_2}{(2\pi)^32E_{k_2}}\nonumber\\
&&\times \delta^{(2)}(q_\perp-k_{2\perp})
S_g(p_1,k_J;k_1)\left[S_g(p_1,k_1;k_2)\right.\nonumber\\
&&\left.+S_g(k_J,k_1;k_2)-S_g(p_1,k_J;k_2)\right] \, ,
\end{eqnarray}
where we have to impose the kinematics of the leading NGLs: $k_1$ belongs to the jet associated with $k_J$ and $k_2$ is out of the jet. For the leading contribution, these conditions can be simplified as,
\begin{equation}
(k_1+k_J)^2<k_{1\perp}k_{J\perp}R^2,~~~(k_2+k_J)^2>k_{2\perp}k_{J\perp}R^2\, ,
\end{equation}
where $R$ is the jet radius. These can be further translated into the following ordering of their momenta,
\begin{eqnarray}
&&\frac{k_{1\perp}^{'2}}{z_1^2}=\frac{(k_{1\perp}-z_1k_{J\perp})^2}{z_1^2}<k_{J\perp}^2R^2\ ,\\
&&\frac{k_{2\perp}^{'2}}{z_2^2}=\frac{(k_{2\perp}-z_2k_{J\perp})^2}{z_2^2}>k_{J\perp}^2R^2\ ,
\end{eqnarray}
where $z_{1,2}$ are defined as $z_i=k_i\cdot p_1/k_J\cdot p_1$. In the soft approximation we have $z_2\ll z_1\ll 1$. For convenience, we have also introduced two new momentum variables: $k_{i\perp}'=k_{i\perp}-z_ik_{J\perp}$. These variables represent the transverse momenta relative to the jet momentum which are obtained by subtracting the momentum components along the jet direction. With this notation, the NGL contribution can be evaluated as
\begin{eqnarray}\label{C7}
&&{\cal S}_{\rm NGL}^{(2)}=\frac{C_FC_A}{2}\left(\frac{\alpha_s}{\pi}\right)^2\int\frac{dz_2}{z_2}\frac{dz_1}{z_1}\frac{d^2k_{2\perp}'}{2\pi}\frac{d^2k_{1\perp}'}{2\pi}\nonumber\\
&&~~\times \delta^{(2)}(q_\perp-k_{2\perp})\Theta(z_1-z_2)\Theta\left(\frac{k_{2\perp}^{'2}}{z_2^2k_{J\perp}^2}-R^2\right)
\nonumber\\
&&~~\times \Theta\left(R^2-\frac{k_{1\perp}^{'2}}{z_1^2k_{J\perp}^2}\right)\frac{1}{k_{1\perp}^{'2}}\frac{\frac{2z_2}{z_1}k_{2\perp}'\cdot k_{1\perp}'}{k_{2\perp}^{'2}(k_{2\perp}'-\frac{z_2}{z_1}k_{1\perp}')^2} \, .\label{e33}
\end{eqnarray}
The next step is to average over the azimuthal angle $\phi_{12}$ between $k_{1\perp}'$ and $k_{2\perp}'$, which leads to the following expression:
\begin{equation}
{\cal I}\equiv
\frac{1}{2\pi}\int_0^{2\pi}d\phi_{12}\frac{\frac{2z_2}{z_1}k_{2\perp}'\cdot k_{1\perp}'}{(k_{2\perp}'-\frac{z_2}{z_1}k_{1\perp}')^2}=
\frac{\frac{k_{2\perp}^{'2}}{z_2^2}+\frac{k_{1\perp}^{'2}}{z_1^2}}{|\frac{k_{2\perp}^{'2}}{z_2^2}-\frac{k_{1\perp}^{'2}}{z_1^2}|}-1 \, .
\end{equation}
For the NGL configuration we have
$k_{2\perp}^{'2}/z_2^2>k_{1\perp}^{'2}/z_1^2$. Therefore, the above result can be written as
\begin{equation}
{\cal I} =
\frac{2\frac{k_{1\perp}^{'2}}{z_1^2}}{\frac{k_{2\perp}^{'2}}{z_2^2}-\frac{k_{1\perp}^{'2}}{z_1^2}} \, .
\end{equation}
We can then carry out the integral over $k_{1\perp}^{'2}$, and we find the following expression,
\begin{equation}
\int_0^{R^2z_1^2k_{J\perp}^2}\frac{dk_{1\perp}^{'2}}{k_{1\perp}^{'2}}\,{\cal I}=2\ln\frac{y}{y-R^2} \, ,
\end{equation}
where $y=k_{2\perp}^{'2}/k_{J\perp}^2z_2^2$. We rewrite the $z_2$ integral as an integral over $y$. Before we carry out the $z_2$ integral, there are large logarithms associated with the strong ordering of $z_2\ll z_1$. Therefore, $\int dz_1/z_1\Theta(z_1-z_2)=\ln(1/z_2)$. When converting the $z_2$ integral to a $y$ integral, there will be a logarithm of $\ln(k_{J\perp}^2/k_{2\perp}^{'2})$. A further approximation to obtain the 
leading contribution of the NGLs is to set $k_{2\perp}'\to k_{2\perp}$ in
this logarithm and the overall factor $1/k_{2\perp}^{'2}$. In the end, we obtain the following result for the NGLs at this order
\begin{equation}
{\cal S}_{\rm NGL}^{(2)}=C_F\frac{C_A}{2}\left(\frac{\alpha_s}{\pi}\right)^2\frac{1}{2\pi}\frac{\pi^2}{6}\frac{1}{q_\perp^2}\ln\frac{k_{J\perp}^2}{q_\perp^2} \, ,
\end{equation}
where the $\pi^2/6$ factor comes from the following integral in the small $R$ limit
\begin{equation}
\lim_{R^2\to 0}\int_{R^2}^\infty \frac{dy}{y}\ln\frac{y}{y-R^2}=\frac{\pi^2}{6} \, .
\end{equation}
The above result is for the real gluon radiation. When Fourier transforming into $b_\perp$-space, and adding the virtual contribution, we obtain
\begin{equation}
\tilde{\cal S}_{\rm NGL}^{(2)}=-C_F\frac{C_A}{2}\left(\frac{\alpha_s}{\pi}\right)^2\frac{\pi^2}{24}\ln^2\frac{k_{J\perp}^2b_\perp^2}{c_0^2} \, .
\end{equation}


\begin{thebibliography}{99}

 %\cite{Liu:2018trl}
\bibitem{Liu:2018trl} 
  X.~Liu, F.~Ringer, W.~Vogelsang and F.~Yuan,
  %``Lepton-jet Correlations in Deep Inelastic Scattering at the Electron-Ion Collider,''
  Phys.\ Rev.\ Lett.\  {\bf 122}, no. 19, 192003 (2019)
  doi:10.1103/PhysRevLett.122.192003
  [arXiv:1812.08077 [hep-ph]].
  %%CITATION = doi:10.1103/PhysRevLett.122.192003;%%
  %5 citations counted in INSPIRE as of 03 Oct 2019
  

%
\bibitem{Boer:2011fh}
  D.~Boer {\it et al.},
  %``Gluons and the quark sea at high energies: Distributions, polarization, tomography,''
  arXiv:1108.1713 [nucl-th].

%\cite{AbelleiraFernandez:2012cc}
\bibitem{AbelleiraFernandez:2012cc} 
  J.~L.~Abelleira Fernandez {\it et al.} [LHeC Study Group],
  %``A Large Hadron Electron Collider at CERN: Report on the Physics and Design Concepts for Machine and Detector,''
  J.\ Phys.\ G {\bf 39}, 075001 (2012)
%  doi:10.1088/0954-3899/39/7/075001
  [arXiv:1206.2913 [physics.acc-ph]].
  %%CITATION = doi:10.1088/0954-3899/39/7/075001;%%

    %
\bibitem{Accardi:2012qut}
  A.~Accardi {\it et al.},
  %``Electron Ion Collider: The Next QCD Frontier - Understanding the glue that binds us all,''
  arXiv:1212.1701 [nucl-ex].
  
  
   %\cite{Abramowicz:2017ful}
\bibitem{Abramowicz:2017ful} 
  H.~Abramowicz {\it et al.} [ZEUS Collaboration],
  %``Further studies of isolated photon production with a jet in deep inelastic scattering at HERA,''
  JHEP {\bf 1801}, 032 (2018)
%  doi:10.1007/JHEP01(2018)032
  [arXiv:1712.04273 [hep-ex]].
  %%CITATION = doi:10.1007/JHEP01(2018)032;%%
  
  %\cite{Abramowicz:2012jz}
\bibitem{Abramowicz:2012jz} 
  H.~Abramowicz {\it et al.} [ZEUS Collaboration],
  %``Inclusive-jet photoproduction at HERA and determination of alphas,''
  Nucl.\ Phys.\ B {\bf 864}, 1 (2012)
%  doi:10.1016/j.nuclphysb.2012.06.006
  [arXiv:1205.6153 [hep-ex]].
  %%CITATION = doi:10.1016/j.nuclphysb.2012.06.006;%%
  %44 citations counted in INSPIRE as of 01 Oct 2018
  
  %\cite{Abramowicz:2010ke}
\bibitem{Abramowicz:2010ke} 
  H.~Abramowicz {\it et al.} [ZEUS Collaboration],
  %``Inclusive-jet cross sections in NC DIS at HERA and a comparison of the kT, anti-kT and SIScone jet algorithms,''
  Phys.\ Lett.\ B {\bf 691}, 127 (2010)
%  doi:10.1016/j.physletb.2010.06.015
  [arXiv:1003.2923 [hep-ex]].
  %%CITATION = doi:10.1016/j.physletb.2010.06.015;%%
  %36 citations counted in INSPIRE as of 01 Oct 2018
  
\bibitem{Amilkar}
 A.~Quintero, EIC 2019 Users Group Annual Meeting, Paris, June 2019; and private communications.
 %

  
  %\cite{Abazov:2004hm}
\bibitem{Abazov:2004hm} 
  V.~M.~Abazov {\it et al.} [D0 Collaboration],
  %``Measurement of dijet azimuthal decorrelations at central rapidities in $p\bar{p}$ collisions at $\sqrt{s} = 1.96$ TeV,''
  Phys.\ Rev.\ Lett.\  {\bf 94}, 221801 (2005)
%  doi:10.1103/PhysRevLett.94.221801
  [hep-ex/0409040].
  %%CITATION = doi:10.1103/PhysRevLett.94.221801;%%
  %150 citations counted in INSPIRE as of 30 Sep 2018
  
  %\cite{Khachatryan:2011zj}
\bibitem{Khachatryan:2011zj} 
  V.~Khachatryan {\it et al.} [CMS Collaboration],
  %``Dijet Azimuthal Decorrelations in $pp$ Collisions at $\sqrt{s} = 7$~TeV,''
  Phys.\ Rev.\ Lett.\  {\bf 106}, 122003 (2011)
%  doi:10.1103/PhysRevLett.106.122003
  [arXiv:1101.5029 [hep-ex]].
  %%CITATION = doi:10.1103/PhysRevLett.106.122003;%%
  %71 citations counted in INSPIRE as of 30 Sep 2018
  
  %\cite{daCosta:2011ni}
\bibitem{daCosta:2011ni} 
  G.~Aad {\it et al.} [ATLAS Collaboration],
  %``Measurement of Dijet Azimuthal Decorrelations in $pp$ Collisions at $\sqrt{s}=7$ TeV,''
  Phys.\ Rev.\ Lett.\  {\bf 106}, 172002 (2011)
%  doi:10.1103/PhysRevLett.106.172002
  [arXiv:1102.2696 [hep-ex]].
  %%CITATION = doi:10.1103/PhysRevLett.106.172002;%%
  %88 citations counted in INSPIRE as of 30 Sep 2018


%\cite{Adamczyk:2013jei}
\bibitem{Adamczyk:2013jei} 
  L.~Adamczyk {\it et al.} [STAR Collaboration],
  %``Jet-Hadron Correlations in $\sqrt{s_{NN}} = 200$ GeV $p+p$ and Central $Au+Au$ Collisions,''
  Phys.\ Rev.\ Lett.\  {\bf 112}, no. 12, 122301 (2014)
%  doi:10.1103/PhysRevLett.112.122301
  [arXiv:1302.6184 [nucl-ex]].
  %%CITATION = doi:10.1103/PhysRevLett.112.122301;%%
  %54 citations counted in INSPIRE as of 17 Dec 2018

%\cite{Adamczyk:2017yhe}
\bibitem{Adamczyk:2017yhe} 
  L.~Adamczyk {\it et al.} [STAR Collaboration],
  %``Measurements of jet quenching with semi-inclusive hadron+jet distributions in Au+Au collisions at $\sqrt{s_{NN}}$ = 200 GeV,''
  Phys.\ Rev.\ C {\bf 96}, no. 2, 024905 (2017)
%  doi:10.1103/PhysRevC.96.024905
  [arXiv:1702.01108 [nucl-ex]].
  %%CITATION = doi:10.1103/PhysRevC.96.024905;%%
  %27 citations counted in INSPIRE as of 17 Dec 2018
 
  %\cite{Collins:1981uk}
\bibitem{Collins:1981uk} 
  J.~C.~Collins and D.~E.~Soper,
  %``Back-To-Back Jets in QCD,''
  Nucl.\ Phys.\ B {\bf 193}, 381 (1981)
  Erratum: [Nucl.\ Phys.\ B {\bf 213}, 545 (1983)].
%  doi:10.1016/0550-3213(81)90339-4
  %%CITATION = doi:10.1016/0550-3213(81)90339-4;%%
  %1087 citations counted in INSPIRE as of 13 Sep 2018
  
  %\cite{Collins:1981uw}
\bibitem{Collins:1981uw} 
  J.~C.~Collins and D.~E.~Soper,
  %``Parton Distribution and Decay Functions,''
  Nucl.\ Phys.\ B {\bf 194}, 445 (1982).
%  doi:10.1016/0550-3213(82)90021-9
  %%CITATION = doi:10.1016/0550-3213(82)90021-9;%%
  %915 citations counted in INSPIRE as of 13 Sep 2018
  
 \bibitem{Collins:1984kg}
  J.~C.~Collins, D.~E.~Soper and G.~F.~Sterman,
  %``Transverse Momentum Distribution in Drell-Yan Pair and W and Z Boson Production,''
  Nucl.\ Phys.\ B {\bf 250}, 199 (1985).
  %%CITATION = NUPHA,B250,199;%%
  
  %\cite{Ji:2004wu}
\bibitem{Ji:2004wu} 
  X.~Ji, J.~P.~Ma and F.~Yuan,
  %``QCD factorization for semi-inclusive deep-inelastic scattering at low transverse momentum,''
  Phys.\ Rev.\ D {\bf 71}, 034005 (2005)
%  doi:10.1103/PhysRevD.71.034005
  [hep-ph/0404183].
  %%CITATION = doi:10.1103/PhysRevD.71.034005;%%
  %523 citations counted in INSPIRE as of 01 Oct 2018
  
  %\cite{GarciaEchevarria:2011rb}
\bibitem{GarciaEchevarria:2011rb} 
  M.~G.~Echevarria, A.~Idilbi and I.~Scimemi,
  %``Factorization Theorem For Drell-Yan At Low q_T And Transverse Momentum Distributions On-The-Light-Cone,''
  JHEP {\bf 1207}, 002 (2012)
%  doi:10.1007/JHEP07(2012)002
  [arXiv:1111.4996 [hep-ph]].
  %%CITATION = doi:10.1007/JHEP07(2012)002;%%
  %203 citations counted in INSPIRE as of 17 Dec 2018

%\cite{Chiu:2012ir}
\bibitem{Chiu:2012ir}
J.~Y.~Chiu, A.~Jain, D.~Neill and I.~Z.~Rothstein,
%``A Formalism for the Systematic Treatment of Rapidity Logarithms in Quantum Field Theory,''
JHEP \textbf{05}, 084 (2012)
doi:10.1007/JHEP05(2012)084
[arXiv:1202.0814 [hep-ph]].
%268 citations counted in INSPIRE as of 15 Jul 2020
  
  %\cite{Collins:2011zzd}
\bibitem{Collins:2011zzd} 
  J.~Collins,
  ``Foundations of perturbative QCD,''
  Camb.\ Monogr.\ Part.\ Phys.\ Nucl.\ Phys.\ Cosmol.\  {\bf 32}, 1 (2011).
  %%CITATION = CMPCE,32,1;%%
  %323 citations counted in INSPIRE as of 01 Oct 2018
  
  %\cite{Mulders:1995dh}
\bibitem{Mulders:1995dh} 
  P.~J.~Mulders and R.~D.~Tangerman,
  %``The Complete tree level result up to order 1/Q for polarized deep inelastic leptoproduction,''
  Nucl.\ Phys.\ B {\bf 461}, 197 (1996)
  Erratum: [Nucl.\ Phys.\ B {\bf 484}, 538 (1997)]
%  doi:10.1016/S0550-3213(96)00648-7, 10.1016/0550-3213(95)00632-X
  [hep-ph/9510301].
  %%CITATION = doi:10.1016/S0550-3213(96)00648-7, 10.1016/0550-3213(95)00632-X;%%
  %778 citations counted in INSPIRE as of 23 Oct 2018
 
 %\cite{Boer:1997nt}
\bibitem{Boer:1997nt} 
  D.~Boer and P.~J.~Mulders,
  %``Time reversal odd distribution functions in leptoproduction,''
  Phys.\ Rev.\ D {\bf 57}, 5780 (1998)
%  doi:10.1103/PhysRevD.57.5780
  [hep-ph/9711485].
  %%CITATION = doi:10.1103/PhysRevD.57.5780;%%
  %789 citations counted in INSPIRE as of 23 Oct 2018
  
  %\cite{Bacchetta:2006tn}
\bibitem{Bacchetta:2006tn} 
  A.~Bacchetta, M.~Diehl, K.~Goeke, A.~Metz, P.~J.~Mulders and M.~Schlegel,
  %``Semi-inclusive deep inelastic scattering at small transverse momentum,''
  JHEP {\bf 0702}, 093 (2007)
%  doi:10.1088/1126-6708/2007/02/093
  [hep-ph/0611265].
  %%CITATION = doi:10.1088/1126-6708/2007/02/093;%%
  %500 citations counted in INSPIRE as of 23 Oct 2018
  
  %\cite{Gutierrez-Reyes:2019vbx}
\bibitem{Gutierrez-Reyes:2019vbx}
D.~Gutierrez-Reyes, I.~Scimemi, W.~J.~Waalewijn and L.~Zoppi,
%``Transverse momentum dependent distributions in $e^+e^-$ and semi-inclusive deep-inelastic scattering using jets,''
JHEP \textbf{10}, 031 (2019)
doi:10.1007/JHEP10(2019)031
[arXiv:1904.04259 [hep-ph]].
%15 citations counted in INSPIRE as of 09 Jul 2020

%\cite{Gutierrez-Reyes:2019msa}
\bibitem{Gutierrez-Reyes:2019msa}
D.~Gutierrez-Reyes, Y.~Makris, V.~Vaidya, I.~Scimemi and L.~Zoppi,
%``Probing Transverse-Momentum Distributions With Groomed Jets,''
JHEP \textbf{08}, 161 (2019)
doi:10.1007/JHEP08(2019)161
[arXiv:1907.05896 [hep-ph]].
%16 citations counted in INSPIRE as of 18 Jul 2020

%\cite{Aschenauer:2019uex}
\bibitem{Aschenauer:2019uex}
E.~C.~Aschenauer, K.~Lee, B.~S.~Page and F.~Ringer,
%``Jet angularities in photoproduction at the Electron-Ion Collider,''
Phys. Rev. D \textbf{101}, no.5, 054028 (2020)
doi:10.1103/PhysRevD.101.054028
[arXiv:1910.11460 [hep-ph]].
%7 citations counted in INSPIRE as of 09 Jul 2020

%\cite{Page:2019gbf}
\bibitem{Page:2019gbf}
B.~S.~Page, X.~Chu and E.~C.~Aschenauer,
%``Experimental Aspects of Jet Physics at a Future EIC,''
Phys. Rev. D \textbf{101}, no.7, 072003 (2020)
doi:10.1103/PhysRevD.101.072003
[arXiv:1911.00657 [hep-ph]].
%5 citations counted in INSPIRE as of 09 Jul 2020

%\cite{Arratia:2019vju}
\bibitem{Arratia:2019vju}
M.~Arratia, Y.~Song, F.~Ringer and B.~Jacak,
%``Jets as precision probes in electron-nucleus collisions at the Electron-Ion Collider,''
Phys. Rev. C \textbf{101}, 065204 (2020)
doi:10.1103/PhysRevC.101.065204
[arXiv:1912.05931 [nucl-ex]].
%10 citations counted in INSPIRE as of 09 Jul 2020

%\cite{Kang:2020xyq}
\bibitem{Kang:2020xyq}
Z.~B.~Kang, K.~Lee and F.~Zhao,
%``Polarized jet fragmentation functions,''
[arXiv:2005.02398 [hep-ph]].
%2 citations counted in INSPIRE as of 09 Jul 2020
%\cite{Arratia:2020ssx}
\bibitem{Arratia:2020ssx}
M.~Arratia, Y.~Makris, D.~Neill, F.~Ringer and N.~Sato,
%``Asymmetric jet clustering in deep-inelastic scattering,''
[arXiv:2006.10751 [hep-ph]].
%1 citations counted in INSPIRE as of 09 Jul 2020

%\cite{Arratia:2020azl}
\bibitem{Arratia:2020azl}
M.~Arratia, Y.~Furletova, T.~J.~Hobbs, F.~Olness and S.~J.~Sekula,
%``Charm jets as a probe for strangeness at the future Electron-Ion Collider,''
[arXiv:2006.12520 [hep-ph]].
%0 citations counted in INSPIRE as of 09 Jul 2020

%\cite{Arratia:2020nxw}
\bibitem{Arratia:2020nxw}
M.~Arratia, Z.~B.~Kang, A.~Prokudin and F.~Ringer,
%``Jet-based measurements of Sivers and Collins asymmetries at the future Electron-Ion Collider,''
[arXiv:2007.07281 [hep-ph]].
%0 citations counted in INSPIRE as of 18 Jul 2020

%\cite{Hinderer:2015hra}
\bibitem{Hinderer:2015hra}
P.~Hinderer, M.~Schlegel and W.~Vogelsang,
%``Single-Inclusive Production of Hadrons and Jets in Lepton-Nucleon Scattering at NLO,''
Phys. Rev. D \textbf{92}, no.1, 014001 (2015)
[arXiv:1505.06415 [hep-ph]].

\bibitem{DAlesio:2017nrd}
U.~D'Alesio, C.~Flore and F.~Murgia,
%``Transverse single-spin asymmetries in $\ell \,p^\uparrow \to h \,X$ within a TMD approach: role of quasi-real photon exchange,''
Phys. Rev. D \textbf{95}, no.9, 094002 (2017)
%doi:10.1103/PhysRevD.95.094002
[arXiv:1701.01148 [hep-ph]].

\bibitem{Boughezal:2018azh}
R.~Boughezal, F.~Petriello and H.~Xing,
%``Inclusive jet production as a probe of polarized parton distribution functions at a future EIC,''
Phys. Rev. D \textbf{98}, no.5, 054031 (2018)
%doi:10.1103/PhysRevD.98.054031
[arXiv:1806.07311 [hep-ph]].

%\cite{Gutierrez-Reyes:2018qez}
\bibitem{Gutierrez-Reyes:2018qez} 
  D.~Gutierrez-Reyes, I.~Scimemi, W.~J.~Waalewijn and L.~Zoppi,
  %``Transverse momentum dependent distributions with jets,''
  Phys.\ Rev.\ Lett.\  {\bf 121}, no. 16, 162001 (2018)
%  doi:10.1103/PhysRevLett.121.162001
  [arXiv:1807.07573 [hep-ph]].
  %%CITATION = doi:10.1103/PhysRevLett.121.162001;%%
  %1 citations counted in INSPIRE as of 21 Nov 2018
  


\bibitem{Borsa:2020ulb}
I.~Borsa, D.~de Florian and I.~Pedron,
%``Jet production in Polarized DIS at NNLO,''
[arXiv:2005.10705 [hep-ph]].

%\cite{Sivers:1989cc}
\bibitem{Sivers:1989cc}
D.~W.~Sivers,
%``Single Spin Production Asymmetries from the Hard Scattering of Point-Like Constituents,''
Phys. Rev. D \textbf{41}, 83 (1990)
doi:10.1103/PhysRevD.41.83
%1120 citations counted in INSPIRE as of 21 Apr 2020


  
  %\cite{Efremov:1981sh}
%\bibitem{Efremov}
\bibitem{et}
  A.~V.~Efremov and O.~V.~Teryaev,
  %``On Spin Effects In Quantum Chromodynamics,''
  Sov.\ J.\ Nucl.\ Phys.\  {\bf 36}, 140 (1982)
  [Yad.\ Fiz.\  {\bf 36}, 242 (1982)];
  %%CITATION = SJNCA,36,140;%%
%\cite{Efremov:1984ip}
%\bibitem{Efremov:1984ip}
  A.~V.~Efremov and O.~V.~Teryaev,
  %``QCD Asymmetry And Polarized Hadron Structure Functions,''
  Phys.\ Lett.\ B {\bf 150}, 383 (1985).
  %%CITATION = PHLTA,B150,383;%%

%\cite{Qiu:pp}
\bibitem{qiusterman}
J.~Qiu and G.~Sterman,
%``Single Transverse Spin Asymmetries,''
Phys.\ Rev.\ Lett.\  {\bf 67}, 2264 (1991);
%%CITATION = PRLTA,67,2264;%%
%\cite{Qiu:1991wg}
%\bibitem{Qiu:1991wg}
%  J.~w.~Qiu and G.~Sterman,
  %``Single transverse spin asymmetries in direct photon production,''
  Nucl.\ Phys.\ B {\bf 378}, 52 (1992);
  %%CITATION = NUPHA,B378,52;%%
%\cite{Qiu:1998ia}
%\bibitem{Qiu:1998ia}
%J.~w.~Qiu and G.~Sterman,
%``Single transverse-spin asymmetries in hadronic pion production,''
Phys.\ Rev.\ D {\bf 59}, 014004 (1999).
%%CITATION = HEP-PH 9806356;%%


\bibitem{Banfi:2008qs} 
  A.~Banfi, M.~Dasgupta and Y.~Delenda,
  %``Azimuthal decorrelations between QCD jets at all orders,''
  Phys.\ Lett.\ B {\bf 665}, 86 (2008)
%  doi:10.1016/j.physletb.2008.05.065
  [arXiv:0804.3786 [hep-ph]].

%\cite{Mueller:2013wwa}
\bibitem{Mueller:2013wwa} 
  A.~H.~Mueller, B.~W.~Xiao and F.~Yuan,
  %``Sudakov double logarithms resummation in hard processes in the small-x saturation formalism,''
  Phys.\ Rev.\ D {\bf 88}, no. 11, 114010 (2013)
%  doi:10.1103/PhysRevD.88.114010
  [arXiv:1308.2993 [hep-ph]].
 
%\cite{Sun:2014gfa}
\bibitem{Sun:2014gfa} 
  P.~Sun, C.-P.~Yuan and F.~Yuan,
  %``Soft Gluon Resummations in Dijet Azimuthal Angular Correlations in Hadronic Collisions,''
  Phys.\ Rev.\ Lett.\  {\bf 113}, no. 23, 232001 (2014)
%  doi:10.1103/PhysRevLett.113.232001
  [arXiv:1405.1105 [hep-ph]].
  %%CITATION = doi:10.1103/PhysRevLett.113.232001;%%
  %21 citations counted in INSPIRE as of 01 Oct 2018

%\cite{Sun:2015doa}
\bibitem{Sun:2015doa} 
  P.~Sun, C.-P.~Yuan and F.~Yuan,
  %``Transverse Momentum Resummation for Dijet Correlation in Hadronic Collisions,''
  Phys.\ Rev.\ D {\bf 92}, no. 9, 094007 (2015)
%  doi:10.1103/PhysRevD.92.094007
  [arXiv:1506.06170 [hep-ph]].
  %%CITATION = doi:10.1103/PhysRevD.92.094007;%%
  %15 citations counted in INSPIRE as of 01 Oct 2018
  
  %\cite{Sun:2016kkh}
\bibitem{Sun:2016kkh} 
  P.~Sun, J.~Isaacson, C.-P.~Yuan and F.~Yuan,
  %``Resummation of High Order Corrections in Higgs Boson Plus Jet Production at the LHC,''
  Phys.\ Lett.\ B {\bf 769}, 57 (2017)
  doi:10.1016/j.physletb.2017.02.037
  [arXiv:1602.08133 [hep-ph]].
  %%CITATION = doi:10.1016/j.physletb.2017.02.037;%%
  %11 citations counted in INSPIRE as of 06 Oct 2019
  
  %\cite{Sun:2018icb}
\bibitem{Sun:2018icb} 
  P.~Sun, B.~Yan, C.-P.~Yuan and F.~Yuan,
  %``Resummation of High Order Corrections in $Z$ Boson Plus Jet Production at the LHC,''
  Phys.\ Rev.\ D {\bf 100}, no. 5, 054032 (2019)
  doi:10.1103/PhysRevD.100.054032
  [arXiv:1810.03804 [hep-ph]].
  %%CITATION = doi:10.1103/PhysRevD.100.054032;%%
  %5 citations counted in INSPIRE as of 06 Oct 2019
  
  %\cite{Sun:2018beb}
\bibitem{Sun:2018beb} 
  P.~Sun, C.-P.~Yuan and F.~Yuan,
  %``Soft Gluon Resummation in Higgs Boson Plus Two Jet Production at the LHC,''
  doi:10.1016/j.physletb.2019.134852
  arXiv:1802.02980 [hep-ph].
  %%CITATION = doi:10.1016/j.physletb.2019.134852;%%
  %3 citations counted in INSPIRE as of 06 Oct 2019
  
    %\cite{Mukherjee:2012uz}
\bibitem{Mukherjee:2012uz} 
  A.~Mukherjee and W.~Vogelsang,
  %``Jet production in (un)polarized pp collisions: dependence on jet algorithm,''
  Phys.\ Rev.\ D {\bf 86}, 094009 (2012)
%  doi:10.1103/PhysRevD.86.094009
  [arXiv:1209.1785 [hep-ph]].
  %%CITATION = doi:10.1103/PhysRevD.86.094009;%%
  %52 citations counted in INSPIRE as of 02 Oct 2018
  
  %\cite{Cacciari:2008gp}
\bibitem{Cacciari:2008gp} 
  M.~Cacciari, G.~P.~Salam and G.~Soyez,
  %``The anti-$k_t$ jet clustering algorithm,''
  JHEP {\bf 0804}, 063 (2008)
%  doi:10.1088/1126-6708/2008/04/063
  [arXiv:0802.1189 [hep-ph]].
  %%CITATION = doi:10.1088/1126-6708/2008/04/063;%%
  %5559 citations counted in INSPIRE as of 06 Nov 2018


  
    %\cite{Boer:2003tx}
\bibitem{Boer:2003tx} 
  D.~Boer and W.~Vogelsang,
  %``Asymmetric jet correlations in p p uparrow scattering,''
  Phys.\ Rev.\ D {\bf 69}, 094025 (2004)
%  doi:10.1103/PhysRevD.69.094025
  [hep-ph/0312320].
  %%CITATION = doi:10.1103/PhysRevD.69.094025;%%
  %107 citations counted in INSPIRE as of 29 Sep 2018

%\cite{Qiu:2007ey}
\bibitem{Qiu:2007ey} 
  J.~W.~Qiu, W.~Vogelsang and F.~Yuan,
  %``Single Transverse-Spin Asymmetry in Hadronic Dijet Production,''
  Phys.\ Rev.\ D {\bf 76}, 074029 (2007)
%  doi:10.1103/PhysRevD.76.074029
  [arXiv:0706.1196 [hep-ph]].
  %%CITATION = doi:10.1103/PhysRevD.76.074029;%%
  %47 citations counted in INSPIRE as of 30 Sep 2018
  
    %\cite{Bomhof:2007su}
\bibitem{Bomhof:2007su} 
  C.~J.~Bomhof, P.~J.~Mulders, W.~Vogelsang and F.~Yuan,
  %``Single-Transverse Spin Asymmetry in Dijet Correlations at Hadron Colliders,''
  Phys.\ Rev.\ D {\bf 75}, 074019 (2007)
%  doi:10.1103/PhysRevD.75.074019
  [hep-ph/0701277 [hep-ph]].
  %%CITATION = doi:10.1103/PhysRevD.75.074019;%%
  %43 citations counted in INSPIRE as of 23 Oct 2018

  %\cite{Collins:2007nk}
\bibitem{Collins:2007nk} 
  J.~Collins and J.~W.~Qiu,
  %``$k_{T}$ factorization is violated in production of high-transverse-momentum particles in hadron-hadron collisions,''
  Phys.\ Rev.\ D {\bf 75}, 114014 (2007)
%  doi:10.1103/PhysRevD.75.114014
  [arXiv:0705.2141 [hep-ph]].
  %%CITATION = doi:10.1103/PhysRevD.75.114014;%%
  %191 citations counted in INSPIRE as of 30 Sep 2018
  
  %\cite{Rogers:2010dm}
\bibitem{Rogers:2010dm} 
  T.~C.~Rogers and P.~J.~Mulders,
  %``No Generalized TMD-Factorization in Hadro-Production of High Transverse Momentum Hadrons,''
  Phys.\ Rev.\ D {\bf 81}, 094006 (2010)
%  doi:10.1103/PhysRevD.81.094006
  [arXiv:1001.2977 [hep-ph]].
  %%CITATION = doi:10.1103/PhysRevD.81.094006;%%
  %192 citations counted in INSPIRE as of 30 Sep 2018
 
\bibitem{Bacchetta:2005rm} 
  A.~Bacchetta, C.~J.~Bomhof, P.~J.~Mulders and F.~Pijlman,
  %``Single spin asymmetries in hadron-hadron collisions,''
  Phys.\ Rev.\ D {\bf 72}, 034030 (2005)
%  doi:10.1103/PhysRevD.72.034030
  [hep-ph/0505268].
    
  %\cite{Vogelsang:2007jk}
\bibitem{Vogelsang:2007jk} 
  W.~Vogelsang and F.~Yuan,
  %``Hadronic Dijet Imbalance and Transverse-Momentum Dependent Parton Distributions,''
  Phys.\ Rev.\ D {\bf 76}, 094013 (2007)
%  doi:10.1103/PhysRevD.76.094013
  [arXiv:0708.4398 [hep-ph]].
  %%CITATION = doi:10.1103/PhysRevD.76.094013;%%
  %67 citations counted in INSPIRE as of 30 Sep 2018
  
  %\cite{Catani:2011st}
\bibitem{Catani:2011st} 
  S.~Catani, D.~de Florian and G.~Rodrigo,
  %``Space-like (versus time-like) collinear limits in QCD: Is factorization violated?,''
  JHEP {\bf 1207}, 026 (2012)
%  doi:10.1007/JHEP07(2012)026
  [arXiv:1112.4405 [hep-ph]].
  %%CITATION = doi:10.1007/JHEP07(2012)026;%%
  %68 citations counted in INSPIRE as of 30 Sep 2018
  
  %\cite{Mitov:2012gt}
\bibitem{Mitov:2012gt} 
  A.~Mitov and G.~Sterman,
  %``Final state interactions in single- and multi-particle inclusive cross sections for hadronic collisions,''
  Phys.\ Rev.\ D {\bf 86}, 114038 (2012)
%  doi:10.1103/PhysRevD.86.114038
  [arXiv:1209.5798 [hep-ph]].
  %%CITATION = doi:10.1103/PhysRevD.86.114038;%%
  %12 citations counted in INSPIRE as of 30 Sep 2018
  
  %\cite{Schwartz:2017nmr}
\bibitem{Schwartz:2017nmr} 
  M.~D.~Schwartz, K.~Yan and H.~X.~Zhu,
  %``Collinear factorization violation and effective field theory,''
  Phys.\ Rev.\ D {\bf 96}, no. 5, 056005 (2017)
%  doi:10.1103/PhysRevD.96.056005
  [arXiv:1703.08572 [hep-ph]].
  %%CITATION = doi:10.1103/PhysRevD.96.056005;%%
  %9 citations counted in INSPIRE as of 30 Sep 2018
  %\cite{Schwartz:2018obd}
  
\bibitem{Schwartz:2018obd} 
  M.~D.~Schwartz, K.~Yan and H.~X.~Zhu,
  %``Factorization Violation and Scale Invariance,''
  Phys.\ Rev.\ D {\bf 97}, no. 9, 096017 (2018)
%  doi:10.1103/PhysRevD.97.096017
  [arXiv:1801.01138 [hep-ph]].
  %%CITATION = doi:10.1103/PhysRevD.97.096017;%%
  %4 citations counted in INSPIRE as of 30 Sep 2018

%\cite{Catani:2013tia}
\bibitem{Catani:2013tia} 
  S.~Catani, L.~Cieri, D.~de Florian, G.~Ferrera and M.~Grazzini,
  %``Universality of transverse-momentum resummation and hard factors at the NNLO,''
  Nucl.\ Phys.\ B {\bf 881}, 414 (2014)
%  doi:10.1016/j.nuclphysb.2014.02.011
  [arXiv:1311.1654 [hep-ph]].
  %%CITATION = doi:10.1016/j.nuclphysb.2014.02.011;%%
  %91 citations counted in INSPIRE as of 23 Oct 2018
%\cite{Prokudin:2015ysa}

%\cite{Rothstein:2016bsq}
\bibitem{Rothstein:2016bsq}
I.~Z.~Rothstein and I.~W.~Stewart,
%``An Effective Field Theory for Forward Scattering and Factorization Violation,''
JHEP \textbf{08}, 025 (2016)
doi:10.1007/JHEP08(2016)025
[arXiv:1601.04695 [hep-ph]].
%100 citations counted in INSPIRE as of 18 Jul 2020

  %\cite{Sun:2013hua}
\bibitem{Sun:2013hua} 
  P.~Sun and F.~Yuan,
  %``Transverse momentum dependent evolution: Matching semi-inclusive deep inelastic scattering processes to Drell-Yan and W/Z boson production,''
  Phys.\ Rev.\ D {\bf 88}, no. 11, 114012 (2013)
%  doi:10.1103/PhysRevD.88.114012
  [arXiv:1308.5003 [hep-ph]].
  %%CITATION = doi:10.1103/PhysRevD.88.114012;%%
  %85 citations counted in INSPIRE as of 29 Sep 2018
  
 

%\cite{Ji:2006br}
\bibitem{Ji:2006br} 
  X.~Ji, J.~W.~Qiu, W.~Vogelsang and F.~Yuan,
  %``Single-transverse spin asymmetry in semi-inclusive deep inelastic scattering,''
  Phys.\ Lett.\ B {\bf 638}, 178 (2006)
%  doi:10.1016/j.physletb.2006.05.044
  [hep-ph/0604128].
  %%CITATION = doi:10.1016/j.physletb.2006.05.044;%%
  %120 citations counted in INSPIRE as of 03 Oct 2018
  
  %\cite{Koike:2007dg}
\bibitem{Koike:2007dg} 
  Y.~Koike, W.~Vogelsang and F.~Yuan,
  %``On the Relation Between Mechanisms for Single-Transverse-Spin Asymmetries,''
  Phys.\ Lett.\ B {\bf 659}, 878 (2008)
%  doi:10.1016/j.physletb.2007.11.096
  [arXiv:0711.0636 [hep-ph]].
  %%CITATION = doi:10.1016/j.physletb.2007.11.096;%%
  %94 citations counted in INSPIRE as of 03 Oct 2018

    %\cite{Braun:2009mi}
\bibitem{Braun:2009mi} 
  V.~M.~Braun, A.~N.~Manashov and B.~Pirnay,
  %``Scale dependence of twist-three contributions to single spin asymmetries,''
  Phys.\ Rev.\ D {\bf 80}, 114002 (2009)
  Erratum: [Phys.\ Rev.\ D {\bf 86}, 119902 (2012)]
%  doi:10.1103/PhysRevD.80.114002, 10.1103/PhysRevD.86.119902
  [arXiv:0909.3410 [hep-ph]].
  %%CITATION = doi:10.1103/PhysRevD.80.114002, 10.1103/PhysRevD.86.119902;%%
  %94 citations counted in INSPIRE as of 26 Oct 2018
  
  %\cite{Kang:2008ey}
\bibitem{Kang:2008ey} 
  Z.~B.~Kang and J.~W.~Qiu,
  %``Evolution of twist-3 multi-parton correlation functions relevant to single transverse-spin asymmetry,''
  Phys.\ Rev.\ D {\bf 79}, 016003 (2009)
%  doi:10.1103/PhysRevD.79.016003
  [arXiv:0811.3101 [hep-ph]].
  %%CITATION = doi:10.1103/PhysRevD.79.016003;%%
  %91 citations counted in INSPIRE as of 26 Oct 2018
  
  
  %\cite{Vogelsang:2009pj}
\bibitem{Vogelsang:2009pj} 
  W.~Vogelsang and F.~Yuan,
  %``Next-to-leading Order Calculation of the Single Transverse Spin Asymmetry in the Drell-Yan Process,''
  Phys.\ Rev.\ D {\bf 79}, 094010 (2009)
%  doi:10.1103/PhysRevD.79.094010
  [arXiv:0904.0410 [hep-ph]].
  %%CITATION = doi:10.1103/PhysRevD.79.094010;%%
  %79 citations counted in INSPIRE as of 26 Oct 2018
  
  %\cite{Zhou:2008mz}
\bibitem{Zhou:2008mz} 
  J.~Zhou, F.~Yuan and Z.~T.~Liang,
  %``QCD Evolution of the Transverse Momentum Dependent Correlations,''
  Phys.\ Rev.\ D {\bf 79}, 114022 (2009)
%  doi:10.1103/PhysRevD.79.114022
  [arXiv:0812.4484 [hep-ph]].
  %%CITATION = doi:10.1103/PhysRevD.79.114022;%%
  %65 citations counted in INSPIRE as of 26 Oct 2018
  
  %\cite{Schafer:2012ra}
\bibitem{Schafer:2012ra} 
  A.~Schafer and J.~Zhou,
  %``A Note on the scale evolution of the ETQS function $T_F(x,x)$,''
  Phys.\ Rev.\ D {\bf 85}, 117501 (2012)
%  doi:10.1103/PhysRevD.85.117501
  [arXiv:1203.5293 [hep-ph]].
  %%CITATION = doi:10.1103/PhysRevD.85.117501;%%
  %27 citations counted in INSPIRE as of 26 Oct 2018
  
  %\cite{Scimemi:2019gge}
\bibitem{Scimemi:2019gge} 
  I.~Scimemi, A.~Tarasov and A.~Vladimirov,
  %``Collinear matching for Sivers function at next-to-leading order,''
  JHEP {\bf 1905}, 125 (2019)
  doi:10.1007/JHEP05(2019)125
  [arXiv:1901.04519 [hep-ph]].
  %%CITATION = doi:10.1007/JHEP05(2019)125;%%
  %6 citations counted in INSPIRE as of 03 Oct 2019
 
 %\cite{Dasgupta:2001sh}
\bibitem{Dasgupta:2001sh} 
  M.~Dasgupta and G.~P.~Salam,
  %``Resummation of nonglobal QCD observables,''
  Phys.\ Lett.\ B {\bf 512}, 323 (2001)
%  doi:10.1016/S0370-2693(01)00725-0
  [hep-ph/0104277].
  %%CITATION = doi:10.1016/S0370-2693(01)00725-0;%%
  %289 citations counted in INSPIRE as of 01 Aug 2018
  
  %\cite{Dasgupta:2002bw}
\bibitem{Dasgupta:2002bw} 
  M.~Dasgupta and G.~P.~Salam,
  %``Accounting for coherence in interjet E(t) flow: A Case study,''
  JHEP {\bf 0203}, 017 (2002)
%  doi:10.1088/1126-6708/2002/03/017
  [hep-ph/0203009].
  %%CITATION = doi:10.1088/1126-6708/2002/03/017;%%
  %143 citations counted in INSPIRE as of 01 Aug 2018

   %\cite{Banfi:2003jj}
\bibitem{Banfi:2003jj} 
  A.~Banfi and M.~Dasgupta,
  %``Dijet rates with symmetric E(t) cuts,''
  JHEP {\bf 0401}, 027 (2004)
%  doi:10.1088/1126-6708/2004/01/027
  [hep-ph/0312108].
  %%CITATION = doi:10.1088/1126-6708/2004/01/027;%%
  %49 citations counted in INSPIRE as of 01 Aug 2018
  %
  
%\cite{Chien:2019gyf}
\bibitem{Chien:2019gyf}
Y.~T.~Chien, D.~Y.~Shao and B.~Wu,
%``Resummation of Boson-Jet Correlation at Hadron Colliders,''
JHEP \textbf{11}, 025 (2019)
doi:10.1007/JHEP11(2019)025
[arXiv:1905.01335 [hep-ph]].
%9 citations counted in INSPIRE as of 19 Jul 2020
  
 \bibitem{Prokudin:2015ysa} 
  A.~Prokudin, P.~Sun and F.~Yuan,
  %``Scheme dependence and transverse momentum distribution interpretation of Collins–Soper–Sterman resummation,''
  Phys.\ Lett.\ B {\bf 750}, 533 (2015)
%  doi:10.1016/j.physletb.2015.09.064
  [arXiv:1505.05588 [hep-ph]].
  %%CITATION = doi:10.1016/j.physletb.2015.09.064;%%
  %17 citations counted in INSPIRE as of 23 Oct 2018
  %

%\cite{Su:2014wpa}
\bibitem{Su:2014wpa} 
  P.~Sun, J.~Isaacson, C.-P.~Yuan and F.~Yuan,
  %``Nonperturbative functions for SIDIS and Drell–Yan processes,''
  Int.\ J.\ Mod.\ Phys.\ A {\bf 33}, no. 11, 1841006 (2018)
%  doi:10.1142/S0217751X18410063
  [arXiv:1406.3073 [hep-ph]].
  %%CITATION = doi:10.1142/S0217751X18410063;%%
  %40 citations counted in INSPIRE as of 29 Sep 2018
  
 
 %\cite{McLerran:1993ni}
\bibitem{McLerran:1993ni}
L.~D.~McLerran and R.~Venugopalan,
%``Computing quark and gluon distribution functions for very large nuclei,''
Phys. Rev. D \textbf{49}, 2233-2241 (1994)
doi:10.1103/PhysRevD.49.2233
[arXiv:hep-ph/9309289 [hep-ph]].
%1967 citations counted in INSPIRE as of 13 Jul 2020

%\cite{McLerran:1994vd}
\bibitem{McLerran:1994vd}
L.~D.~McLerran and R.~Venugopalan,
%``Green's functions in the color field of a large nucleus,''
Phys. Rev. D \textbf{50}, 2225-2233 (1994)
doi:10.1103/PhysRevD.50.2225
[arXiv:hep-ph/9402335 [hep-ph]].
%997 citations counted in INSPIRE as of 13 Jul 2020
 
 %\cite{Marquet:2009ca}
\bibitem{Marquet:2009ca}
C.~Marquet, B.~W.~Xiao and F.~Yuan,
%``Semi-inclusive Deep Inelastic Scattering at small x,''
Phys. Lett. B \textbf{682}, 207-211 (2009)
doi:10.1016/j.physletb.2009.10.099
[arXiv:0906.1454 [hep-ph]].
%35 citations counted in INSPIRE as of 13 Jul 2020


\end{thebibliography}
\end{document}